\documentclass[prd,reprint,groupedaddress,nofootinbib]{revtex4-2}

\usepackage{graphicx}% Include figure files
\usepackage{dcolumn}% Align table columns on decimal point
\usepackage{bm}% bold math
\usepackage{mathtools}
\usepackage{amssymb,amsmath}
\usepackage[protrusion=true,expansion=true]{microtype}
\usepackage[dvipsnames]{xcolor}
\usepackage{microtype}
\usepackage{siunitx}
\usepackage{multirow}
\usepackage{dsfont}
\usepackage{slashed}
\usepackage[colorlinks=true,linkcolor=blue,citecolor=blue,urlcolor=blue]{hyperref}

\usepackage{ragged2e}
\newcolumntype{L}[1]{>{\raggedright\arraybackslash}p{#1}}
\newcolumntype{C}[1]{>{\centering\arraybackslash}p{#1}}
\newcolumntype{R}[1]{>{\raggedleft\arraybackslash}p{#1}}
\newcolumntype{J}[1]{>{\justifying\arraybackslash}p{#1}}

\newcommand{\F}{\mathcal{F}}

        \def\mC{\ensuremath{\mathcal{C}}}
        
        \def\mF{\ensuremath{\mathcal{F}}}

\begin{document}

\title{Axial-vector and scalar contributions to hadronic light-by-light scattering}

\author{Gernot Eichmann$^1$}
\email[]{gernot.eichmann@uni-graz.at}
\author{Christian S. Fischer$^{2,3}$}
\email[]{christian.fischer@theo.physik.uni-giessen.de}
\author{Tim Haeuser$^{2}$}
\author{Oliver Regenfelder$^1$}

\affiliation{$^1$Institute of Physics, University of Graz, NAWI Graz, Universitätsplatz 5, 8010 Graz, Austria}
\affiliation{$^2$Institut für Theoretische Physik, Justus-Liebig-Universität Gießen, 35392 Gießen, Germany}
\affiliation{$^3$Helmholtz Forschungsakademie Hessen für FAIR (HFHF), GSI Helmholtzzentrum
für Schwerionenforschung, Campus Gießen, 35392 Gießen, Germany}

\date{\today}

\begin{abstract}
We present results for single axial-vector and scalar meson pole contributions to the hadronic 
light-by-light scattering (HLbL) part of the muon's anomalous magnetic moment. In the dispersive 
approach to these quantities (in narrow width approximation) the central inputs are the
corresponding space-like electromagnetic transition form factors. We determine these 
directly using a functional approach to QCD by Dyson-Schwinger and Bethe-Salpeter equations in
the very same setup we used previously to determine pseudo-scalar meson exchange 
($\pi$, $\eta$ and $\eta'$) as well as meson  ($\pi$ and $K$) box contributions. 
Particular care is taken to preserve gauge invariance and to comply with short distance 
constraints in both the form factors and the HLbL tensor. Our result for the contributions 
from a tower of axial-vector states including short distance constraints is {
$a_\mu^{\text{HLbL}}[\text{AV-tower+SDC}] = 24.8 \,(6.1)  \times 10^{-11}$}.
For the combined contributions from $f_0(980), a_0(980), f_0(1370)$ and $a_0(1450)$ we find 
$a_\mu^{\text{HLbL}}[\text{scalar}] = -1.6 \,(5)  \times 10^{-11}$. 
\end{abstract}

% insert suggested keywords - APS authors don't need to do this
%\keywords{}

\maketitle

\section{Introduction}\label{intro}

\begin{figure}[b]
	\begin{center}
		\includegraphics[scale = 1]{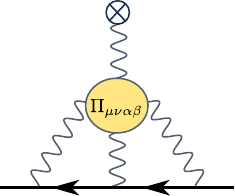} \hfill
		\includegraphics[scale = 1]{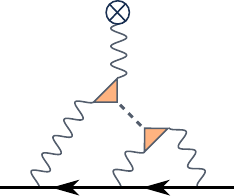}
	\end{center}
	\caption{\textit{Left:} HLbL scattering contribution to $a_\mu$.
		The main ingredient is the hadronic photon four-point function $\Pi_{\mu\nu\alpha\beta}$.
		 \textit{Right:} Meson-exchange part of the HLbL contribution to $a_\mu$ 
		       (without permutations of the photon legs).}
	\label{fig:lbl_contributions}
\end{figure}

The anomalous magnetic moment $a_\mu = \frac{1}{2}(g-2)_\mu$ of the muon is under intense 
scrutiny from both theory and experiment. With updated results from 2019 and 2020 data runs of 
the Fermilab muon $g-2$ experiment, the experimental world average 
$a_\mu (\mbox{exp})=116\,592\,059\,(22) \times 10^{-11}$ \cite{Muong-2:2023cdq} 
has an uncertainty reduced by more than a factor of two  compared
to the earlier BNL experiment \cite{Bennett:2006fi,Roberts:2010cj}. The theoretical error of the 
Standard Model prediction of this quantity \cite{Aoyama:2020ynm,Aoyama:2012wk,Aoyama:2019ryr,Czarnecki:2002nt,
	Gnendiger:2013pva,Davier:2017zfy,Keshavarzi:2018mgv,Colangelo:2018mtw,
	Hoferichter:2019mqg,Davier:2019can,Keshavarzi:2019abf,Hoid:2020xjs,
	Kurz:2014wya,Melnikov:2003xd,Colangelo:2014dfa,Colangelo:2014pva,
	Colangelo:2015ama,Masjuan:2017tvw,Colangelo:2017qdm,Colangelo:2017fiz,
	Hoferichter:2018dmo,Hoferichter:2018kwz,Gerardin:2019vio,Bijnens:2019ghy,
	Colangelo:2019lpu,Colangelo:2019uex,Blum:2019ugy,Colangelo:2014qya} 
is dominated by the error in the hadronic contributions, 
i.e. the hadronic vacuum polarisation (HVP) and hadronic light-by-light scattering (HLbL) 
displayed in Fig.~\ref{fig:lbl_contributions}. 
In order to reduce this theoretical error it is mandatory to improve both.
The size of the HLbL contribution to $a_\mu$ has been estimated in a White Paper of the Muon $g-2$ Theory Initiative,
$a_\mu (\mbox{HLbL})=92 \,(19) \times 10^{-11}$
\cite{Aoyama:2020ynm,Melnikov:2003xd,Masjuan:2017tvw,Colangelo:2017qdm,Colangelo:2017fiz,
	Hoferichter:2018dmo,Hoferichter:2018kwz,Gerardin:2019vio,Bijnens:2019ghy,Colangelo:2019lpu,
	Colangelo:2019uex,Pauk:2014rta,Danilkin:2016hnh,Jegerlehner:2017gek,Knecht:2018sci,
	Eichmann:2019bqf,Roig:2019reh}, 
with an error budget dominated by uncertainties in the contributions from axial-vector meson poles and 
short-distance physics. Contributions from scalar mesons beyond the leading $s$-wave rescattering effects 
also have large uncertainties but are considered to be much smaller in size. In this work we focus  on 
 contributions from intermediate meson states with both axial-vector and scalar quantum numbers with the
goal to contribute to reducing these uncertainties. 

While leading pole contributions with pseudoscalar quantum numbers ($\pi_0, \eta, \eta'$) are well
under control \cite{Knecht:2001qf,Colangelo:2015ama,Nyffeler:2016gnb,Gerardin:2016cqj,
	Masjuan:2017tvw,Hoferichter:2018dmo,Hoferichter:2018kwz,
	Gerardin:2019vio,Eichmann:2019tjk,Raya:2019dnh,Gerardin:2023naa,ExtendedTwistedMass:2023hin}, 
subleading contributions from axial-vectors are still under much scrutiny. Technical 
issues such as the problem of kinematic singularities in the basis of the HLbL tensor have been 
addressed in Ref.~\cite{Hoferichter:2024fsj}, leading to an optimised basis which we will utilize 
in this work. Progress has also been made in identifying necessary ingredients for satisfying 
short-distance constraints (SDC) \cite{Bijnens:2019ghy,Bijnens:2020xnl,Bijnens:2021jqo,Bijnens:2022itw}. These were identified by Melnikov and Vainshtein 
\cite{Melnikov:2003xd} and have  subsequently led to intense debates on the detailed mechanism of 
their realisation in terms of the axial anomaly, single- and double off-shell pseudoscalar and 
axial-vector transition form factors (axTFF), and contributions from infinite towers 
of pseudoscalar and axial-vector states. Recent discussions on this issue can be found in
\cite{Colangelo:2019lpu,Colangelo:2019uex,Melnikov:2019xkq,Leutgeb:2019gbz,Cappiello:2019hwh,Masjuan:2020jsf,
	Ludtke:2020moa,Colangelo:2021nkr,Leutgeb:2021mpu,Leutgeb:2022lqw,Hoferichter:2023tgp,Ludtke:2024ase}.

Another technical complication in the determination of the transition form factors (TFFs) of the axial-vector (AV)
and scalar (S) states is gauge invariance. The general tensor decomposition of the corresponding 
current leads to six (AV) and five (S) different tensor structures, which reduce to five each due to transversality 
of the axial-vector meson. Furthermore, the corresponding form factors of two (AV) and three (S) structures must vanish 
due to gauge invariance. The latter constraint is easily overlooked in {quark} model calculations but is vital to 
guarantee the absence of model artefacts in the remaining physical form factors. 

In this work we  employ a microscopic approach to QCD that satisfies both small and large momentum constraints 
on the TTFs as well as constraints due to gauge invariance. We work with a rainbow-ladder truncation of the Dyson-Schwinger 
and Bethe-Salpeter equations (DSEs and BSEs) of QCD  that has  delivered results for the pseudoscalar pion, 
$\eta$ and $\eta'$ pole contributions to HLbL
\cite{Eichmann:2017wil,Weil:2017knt,Eichmann:2019tjk}
in very good agreement with the data-driven dispersive approach as well as lattice QCD.\footnote{Our framework 
also delivers results 
for the HVP contribution in the ballpark of dispersive and lattice results, although the error of  
roughly 3\% is much too large to make any claims for this quantity~\cite{Goecke:2011pe}.} 
The same is true for the pion box contribution \cite{Eichmann:2019tjk}. Furthermore, our prediction for 
the kaon box contribution \cite{Eichmann:2019bqf} has later been confirmed  by the dispersive
approach~\cite{Stamen:2022uqh}. 

In the next Sec.~\ref{tffs} we detail our calculations for the axial-vector and scalar transition form 
factors and discuss our results. In Sec.~\ref{HLbL} we present corresponding results for
HLbL and we conclude in Sec.~\ref{summary}. We use a Euclidean notation throughout this work; 
see e.g. Appendix A of Ref.~\cite{Eichmann:2016yit} for conventions.

\section{Transition form factors}\label{tffs}

  \subsection{Matrix elements}

\begin{figure}[t]
                    \begin{center}
                    \includegraphics[width=1\columnwidth]{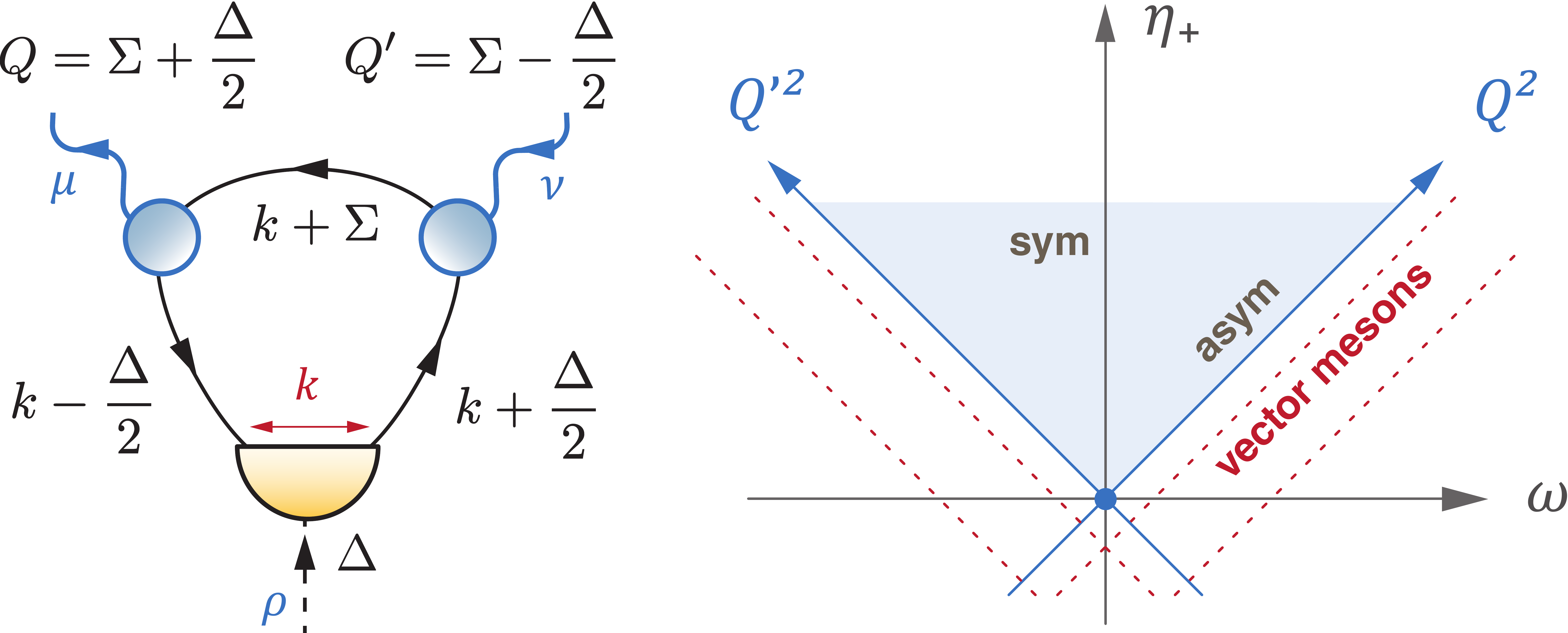}
                    \caption{\textit{Left:} Meson $\to\gamma^{(\ast)}\gamma^{(\ast)}$ transition matrix element. 
                             \textit{Right:} Kinematic domains in $Q^2$ and ${Q'}^2$  including the symmetric and asymmetric limits;
                             the dotted lines show the vector-meson pole locations.
                             }\label{fig:phasespace-1}
                    \end{center}
            \end{figure}

  We consider the axial-vector two-photon transition matrix element $\Lambda_A^{\mu\nu\rho}(Q,Q')$
  and its scalar counterpart $\Lambda_S^{\mu\nu}(Q,Q')$, which we collectively denote by $\Lambda^{\mu\nu(\rho)}_\text{M}(Q,Q')$  with $\text{M}=A, S$.
  Our conventions are shown in Fig.~\ref{fig:phasespace-1}: $Q'$ and $Q$ are the incoming and outgoing photon four-momenta, respectively. 
  They can be expressed through the average photon momentum $\Sigma = (Q+Q')/2$ and the incoming meson momentum
  $\Delta = Q-Q'$, which is onshell ($\Delta^2 = -m_\text{M}^2)$.
   The process then depends on two Lorentz invariants, either $Q^2$ and ${Q'}^2$ or their linear combinations
  \begin{equation}\label{vars-1}
     \eta_+ = \frac{Q^2 + {Q'}^2}{2}\,, \quad
     \omega    =  \frac{Q^2 - {Q'}^2}{2} \,,
  \end{equation}
  while the third variable is fixed by the onshell constraint: 
  \begin{equation}
      \eta_- = Q\cdot Q' = \frac{Q^2 + {Q'}^2 + m_M^2}{2}\,.
  \end{equation}
   The amplitude is Bose-symmetric,  
   \begin{equation} \label{bose}
    \Lambda^{\mu\nu(\rho)}_\text{M}(Q,Q') = \Lambda^{\nu\mu(\rho)}_\text{M}(-Q',-Q)\,,
   \end{equation}   
   and electromagnetic gauge invariance implies transversality in the photon legs:
   \begin{equation}\label{em-gi}
     Q^\mu \Lambda_\text{M}^{\mu\nu(\rho)}(Q,Q') = 0\,, \qquad
     {Q'}^\nu \Lambda_\text{M}^{\mu\nu(\rho)}(Q,Q') = 0\,.
   \end{equation}

   In App.~\ref{sec:basis} we show that the most general tensor decomposition of the axial-vector matrix element
   that implements Bose symmetry and is free of kinematic constraints can be written as
   \begin{equation}\label{amp-final}
    \Lambda_A^{\mu\nu\rho}(Q,Q') = m_A \sum_{i=1}^6 F_i^A(Q^2, {Q'}^2)\,\tau_{i}^{\mu\nu\rho}(Q,Q')\,,
   \end{equation}
   with six dimensionless form factors $F_i^A$  and corresponding dimensionless Lorentz tensors $\tau_i$:
 \begin{equation}\label{taui-A}
 \begin{split} 
    m_A^3\,\tau_1^{\mu\nu\rho} & = \frac{1}{2}\left( \varepsilon^{\mu\rho\alpha}_Q \,t^{\alpha\nu}_{Q'Q'} - t^{\mu\alpha}_{QQ}\,\varepsilon^{\nu\rho\alpha}_{Q'}\right)\,,  \\[1mm]
    m_A^5\,\tau_2^{\mu\nu\rho} & = \frac{\omega}{2} \left( \varepsilon^{\mu\rho\alpha}_Q \,t^{\alpha\nu}_{Q'Q'} + t^{\mu\alpha}_{QQ}\,\varepsilon^{\nu\rho\alpha}_{Q'} \right)\,,   \\[1mm]
    m_A^5\,\tau_3^{\mu\nu\rho} & =  2\omega\,\varepsilon^{\mu\nu}_{QQ'} \Sigma^\rho\,, \\[1mm]
    m_A^3\,\tau_4^{\mu\nu\rho} & =  \varepsilon^{\mu\nu}_{QQ'} \Delta^\rho\,, \\[1mm]
   m_A\,\tau_5^{\mu\nu\rho} &  = \varepsilon^{\mu\nu\rho}_\Sigma\,, \\[1mm]
    m_A^3\,\tau_6^{\mu\nu\rho} & = \omega\,\varepsilon^{\mu\nu\rho}_\Delta\,.
 \end{split}
 \end{equation}  
 Here and in the following we abbreviate
 \begin{equation}
 \begin{array}{rl}
    \varepsilon^{\mu\alpha\beta}_a &\!\!\!= \varepsilon^{\mu\alpha\beta\lambda} a^\lambda\,, \\[2mm]
    \varepsilon^{\mu\nu}_{ab}  &\!\!\!= \varepsilon^{\mu\nu\alpha\beta} a^\alpha b^\beta\,, 
 \end{array}\quad
 \begin{array}{rl}
    t^{\mu\alpha\beta}_a &\!\!\!= \delta^{\mu\beta} a^\alpha - \delta^{\mu\alpha} a^\beta\,, \\[2mm]
    t^{\mu\nu}_{ab} &\!\!\!= a\cdot b \,\delta^{\mu\nu} - b^\mu a^\nu\,,
 \end{array}
 \end{equation}
 where $a$ and $b$ are four-momenta.
 Electromagnetic gauge invariance eliminates the tensors $\tau_5$ and $\tau_6$, which entails that $F_5^A$ and $F_6^A$ must vanish.
 Our calculation satisfies this requirement as long as all ingredients entering in the matrix element
 are calculated consistently from the quark level; however, as discussed in App.~\ref{sec:gi} this is not automatic if one employs model inputs.
 From $a^\mu \,t^{\mu\nu}_{ab} = 0$ and $t^{\mu\nu}_{ab}\,b^\nu = 0$ the transversality of the remaining tensors $\tau_1 \dots \tau_4$ is manifest. 
 Finally, the transversality of the axial-vector meson also eliminates $\tau_4$ (see App.~\ref{sec:basis}).
 
 The $F_i^A$ are Bose-symmetric 
  and therefore even in the variable $\omega$,
 so they can be written as $F_i^A(\eta_+,\omega^2)$. Their kinematic region is shown in the right panel of Fig.~\ref{fig:phasespace-1}:
 The  two asymmetric (singly-virtual) limits correspond to $\omega=\pm \eta_+$ and the symmetric (doubly-virtual) 
 limit implies $\omega=0$.
 In the timelike region, the form factors must have vector-meson poles.
Our form factors are related to the three form factors $\mF_\text{s}$, $\mF_{\text{a}_1}$ and $\mF_{\text{a}_2}$ 
in Refs.~\cite{Hoferichter:2020lap,Hoferichter:2023tgp} by
  \begin{equation}\label{hoferichter-comparison}
     F_1^A = \mF_\text{s}\,, \quad  F_2^A = -\frac{m_A^2}{\omega}\,\mF_{\text{a}_2}\,, \quad  F_3^A = -\frac{m_A^2}{\omega}\,\mF_{\text{a}_1}\,.
 \end{equation}

 The analogous decomposition for the scalar matrix element is given by
   \begin{equation}\label{amp-final-sc}
    \Lambda_S^{\mu\nu}(Q,Q') = m_S \sum_{i=1}^5 F_i^S(Q^2, {Q'}^2)\,\tau_{i}^{\mu\nu}(Q,Q') 
   \end{equation}
  with
 \begin{equation}\label{taui-sc}
 \begin{split} 
    m_S^2\,\tau_1^{\mu\nu} & = - t^{\mu\nu}_{QQ'}\,,  \\[1mm]
    m_S^4\,\tau_2^{\mu\nu} & = t^{\mu\alpha}_{QQ}\,t^{\alpha\nu}_{Q'Q'}   \\[1mm]
           \tau_3^{\mu\nu} & =  \delta^{\mu\nu}\,, \\[1mm]
    m_S^2\,\tau_4^{\mu\nu} & =  Q^\mu {Q'}^\nu\,, \\[1mm]
    m_S^2\,\tau_5^{\mu\nu} &  = Q^\mu Q^\nu + {Q'}^\mu {Q'}^\nu\,.
 \end{split}
 \end{equation}  
 In this case electromagnetic gauge invariance eliminates all tensors except $\tau_1$ and $\tau_2$,
 which leaves two dimensionless form factors $F_{1,2}^S(\eta_+,\omega^2)$.

  \subsection{Microscopic calculation}\label{sec:micro}
 
       As in our previous works for the pseudoscalar transition form factors~\cite{Eichmann:2017wil,Weil:2017knt,Eichmann:2019tjk},
       we employ a rainbow-ladder truncation whose details can be found 
       in Refs.~\cite{Maris:1999bh,Eichmann:2016yit}. For the convenience 
       of the reader we also provide all explicit equations with short explanations in App.~\ref{sec:RL}.   
       
       The microscopic decomposition of the matrix element is then given by (see Fig.~\ref{fig:phasespace-1})
	\begin{equation} \label{mic-decomp}
    \begin{split}
            \Lambda^{\mu\nu(\rho)}_\text{M} &= 2 \,c_M\, \text{Tr} \int \!\! \frac{d^4k}{(2\pi)^4} \,  S(k_+)\,\Gamma^{(\rho)}_\text{M}(k,\Delta)\,S(k_-)  \\
            &\times \Gamma^\mu(k_-,k+\Sigma)\,S(k+\Sigma)\,\Gamma^\nu(k+\Sigma,k_+) \,.
    \end{split}
	\end{equation}
    Here, $k_\pm = k \pm \Delta/2$ are the quark momenta. 
    $S$ denotes the fully dressed quark propagator, $\Gamma^{(\rho)}_\text{M}$ the Bethe-Salpeter amplitude of the
    scalar and axial-vector meson, respectively, and $\Gamma^\mu$ is the fully dressed quark-photon vertex.    
        The factor 2 in front comes from the symmetrization of the photon legs, which yields two identical diagrams,
        and $c_M$ is the colour-flavour trace.
     The rainbow-ladder truncation induces ideal mixing and thus the $a_1$ and $f_1$ are mass degenerate whereas the $f_1'$ is a pure $s\bar{s}$ state.
        Because our meson Bethe-Salpeter amplitudes are normalised to~1 in colour and flavour space,
        the colour trace is $\sqrt{3}$ and the combined colour-flavour factors are
     \begin{equation}\label{colour-flavour}
         c_{a_{0,1}} = \frac{1}{\sqrt{6}}, \quad
         c_{f_{0,1}} = \frac{5}{3\sqrt{6}}, \quad
         c_{f'_{0,1}} = \frac{1}{3\sqrt{3}}\,.
     \end{equation}
     Note  that on the r.h.s. of Eqs.~\eqref{amp-final} and~\eqref{mic-decomp} we suppressed the squared electric charge $e^2$ for brevity.

             \begin{figure}[t]
                    \begin{center}
                    \includegraphics[width=0.9\columnwidth]{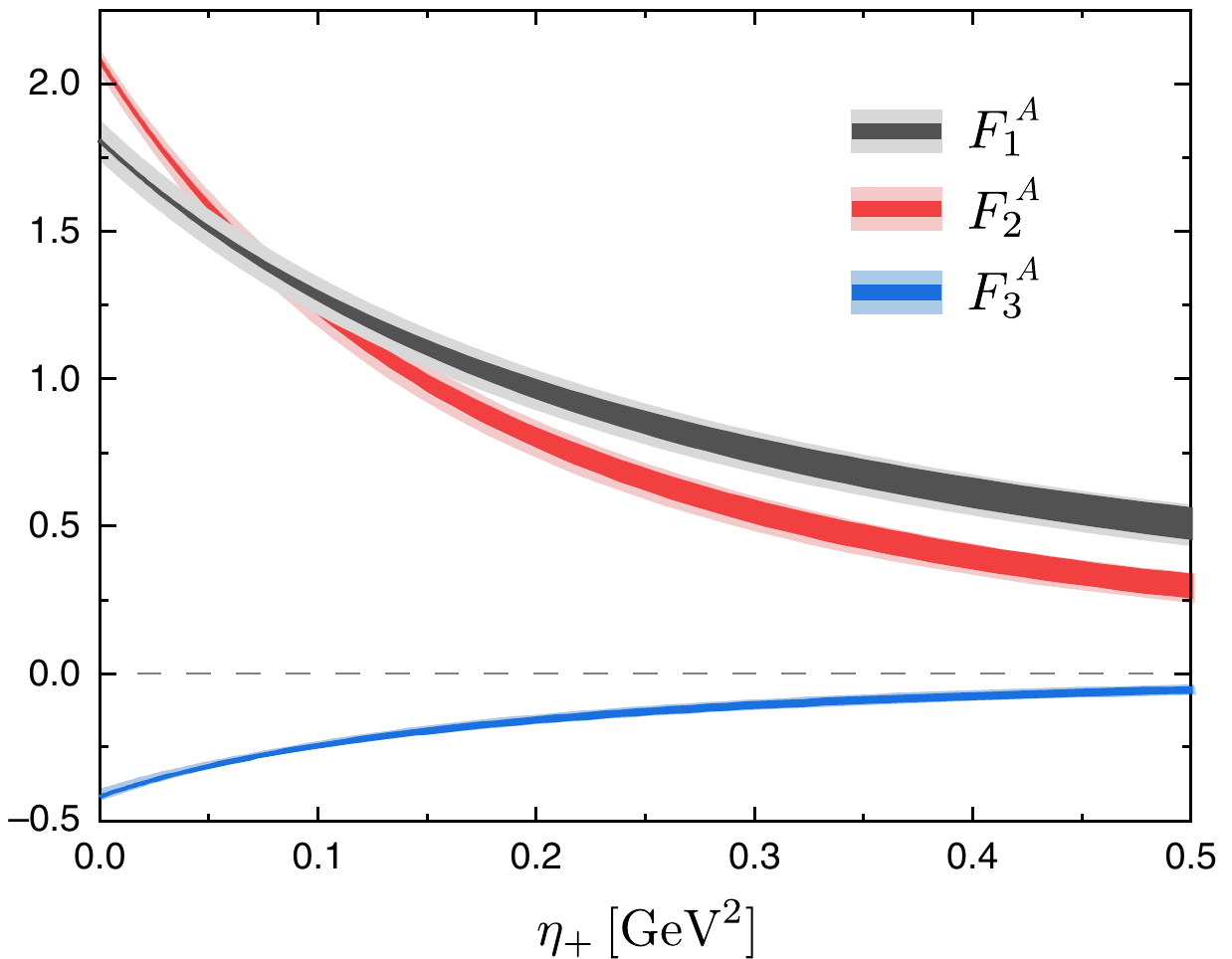}
                    \caption{Axial-vector transition form factors as functions of the average photon momentum $\eta_+ = (Q^2+{Q'}^2)/2$.
                             The darker bands cover the kinematic domain between the symmetric and asymmetric limits.
                             The lighter bands include also the dependence on the shape parameter in the effective interaction.
                             }\label{fig:fax-full}
                    \end{center}
                    \vspace{-2mm}
            \end{figure} 
     
    Eq.~(\ref{mic-decomp}) depends on several nonperturbative ingredients, which we determine from numerical solutions of their DSEs and BSEs.
The renormalised dressed quark propagator
$S^{-1}(p) = Z_f(p^2)\,(i \slashed{p}  + M(p^2))$
involves the wave function $Z_f(p^2)$ and the quark mass function $M(p^2)$, which encodes effects
of dynamical mass generation due to the dynamical breaking of chiral symmetry.
The quark-photon vertex $\Gamma^\mu$ can be decomposed into twelve tensors; 
see e.g. App.~B of Ref.~\cite{Eichmann:2016yit} for details.
Our numerical solution for the vertex
from the inhomogeneous Bethe-Salpeter equation~\cite{Maris:1999bh,Maris:1999ta,Maris:2002mz,Bhagwat:2006pu,Goecke:2010if}
dynamically generates timelike vector-meson poles in its transverse part, so
the underlying physics of vector-meson dominance is
already contained in the form factors without the need for further adjustments.
Finally, the meson amplitudes $\Gamma^{(\rho)}_\text{M}$ 
are determined from their homogeneous BSEs, see App.~\ref{sec:RL}.

In a  rainbow-ladder calculation the masses of ground state pseudoscalar and vector mesons 
come out close to the experimental ones, whereas the axial-vector and scalar meson masses are quite 
far away: $m_A = 0.91$~GeV for  $a_1$/$f_1$ and $m_S = 0.67$~GeV for  $a_0$/$f_0$. Indeed, corrections 
beyond rainbow-ladder are needed to achieve agreement with experiment~\cite{Chang:2011ei,Williams:2015cvx},
and four-body calculations indicate that the $f_0(500)$ is predominantly a four-quark state with only 
a small $q\bar{q}$ component~\cite{Heupel:2012ua,Santowsky:2020pwd,Santowsky:2021ugd}.

{This mass discrepancy has a direct influence on the transition form factors via the various factors of $m_A$ and $m_S$
	showing up in the matrix elements discussed above.} 
We quote in the following the form factors $F_i^{A,S}$ obtained by multiplying the 
rainbow-ladder results $(F_i^{A,S})^\text{RL}$ 
with powers of $(m_\text{M}^\text{exp}/m_\text{M}^\text{RL})$ according to
\begin{equation}\label{resc}
  \sum_i (F_i^{A,S})\,\tau_i = \sum_{i} (F_i^{A,S})^\text{RL}\,\tau_i^\text{RL} \,. 
\end{equation}
Here, the $\tau_i$ are the tensors in Eqs.~\eqref{taui-A} and~\eqref{taui-sc} with  
the experimental $a_1$ and $a_0$ masses, $m_A^\text{exp} = 1.23$ GeV and $m_S^\text{exp} = 0.98$ GeV,
and the $\tau_i^\text{RL}$ are those using the rainbow-ladder masses $m_A$, $m_S$. These results are
presented and discussed in sections \ref{sec:constraints} and \ref{sec:scTFF}.  
{It turns out that the axialvector TFFs treated in this way lead to an acceptable 
value for the equivalent two-photon decay width of the $f_1(1285)$, whereas the scalar two-photon 
decay widths are not well represented. In order to treat all TFFs on the same footing, and to account for 
experimental constraints, we therefore choose a `data-driven' approach by using the known experimental 
two-photon decay widths of the $f_1(1285)$ and the $f_0(980)$, $a_0(980)$, $f_0(1370)$ and $a_0(1450)$
as inputs for the normalisation of the TFFs, which are then employed  in our calculations of the $a_\mu$ contributions
in Sec.~\ref{HLbL} below.}

  \subsection{Axial-vector transition form factors}\label{sec:constraints}

  Our results for the three axial-vector transition form factors $F_i^A(Q^2,{Q'}^2)$ with $i=1,2,3$ are shown in Fig.~\ref{fig:fax-full}.
  The plotted curves correspond to the total contributions %$F_{i,a_1}^A + F_{i,f_1}^A + F_{i,f_1'}^A$ 
  summing over the $a_1$, $f_1$ and $f_1'$ via Eq.~\eqref{colour-flavour}
  under the assumption that their individual form factors are identical.
  The plot covers the full kinematic domain between the symmetric and asymmetric limits (dark bands).
  As in the case of the pion transition form factor~\cite{Eichmann:2017wil,Weil:2017knt}, the angular 
  dependence is very small and collapses into a narrow band in the average photon momentum $\eta_+$.

 The authors of Ref.~\cite{Zanke:2021wiq,Hoferichter:2023tgp} provide parameterisations for the three form factors based on various experimental reactions involving the $f_1(1285)$, namely
 $e^+e^- \to e^+ e^- f_1$, $e^+ e^- \to f_1 \pi^+ \pi^-$ and the radiative decays $f_1 \to \rho\gamma$ and $f_1 \to\phi\gamma$.
 Expressed in terms of our $F_i^A$ using Eq.~\eqref{hoferichter-comparison}, and summing
  again over the $a_1$, $f_1$ and $f_1'$,
 in the limit $Q^2 = {Q'}^2 = 0$  this yields the  constraints\footnote{To be compatible with our conventions
 with positive flavour factors, we switched the sign of the $f_1'$ contribution
  in Ref.~\cite{Hoferichter:2023tgp} since the authors employ a minus sign in their flavour basis.}   
 \begin{equation}\label{ffs-exp}
    \left[ 2.45\,(35) \atop 1.96\,(42)\right], \quad
    \left[ 2.18\,(1.17) \atop 0.42\,(1.49)\right], \quad
    \left[ -0.74\,(84) \atop -0.33\,(84)\right] 
 \end{equation}
 for $F_1^A$, $F_2^A$ and $F_3^A$, respectively,
 where the lower values include an additional constraint from $f_1 \to\phi\gamma$.

Fig.~\ref{fig:fax-full} shows that our form factors are  compatible with  these values:
 $F_1^A$ and $F_2^A$ are large and positive while $F_3^A$ is smaller and negative.
 Our results at $Q^2 = {Q'}^2 = 0$ are 
 \begin{equation}\label{ourFFs-zero}
    F_1^A = 1.80\,(7) , \quad  F_2^A = 2.08\,(5) , \quad F_3^A = -0.41\,(1), 
 \end{equation}
 where the errors come from a variation of the shape parameter $\eta = 1.8 \pm 0.2$ in the effective quark-gluon interaction. 
 The corresponding errors at non-zero momenta $Q^2,{Q'}^2$ are visible in Fig.~\ref{fig:fax-full} as
 light coloured bands. These correspond to similar error bands as in our results of Ref.~\cite{Eichmann:2019tjk}
 for the pseudoscalar TFFs.

{The form factor $F_1^A$ is related to the equivalent two-photon decay width
\begin{align}\label{ax-2gdecaywith}
	\Gamma_{\gamma \gamma} = \frac{\pi \alpha^2}{48}\,m_A |F^A_1(0,0)|^2\,,
\end{align}
whose experimental value for  the $f_1(1285)$ measured by the L3 collaboration is $\Gamma_{\gamma \gamma} = 3.5 (6)(5)$ keV~\cite{L3:2001cyf}.
According to Eq.~\eqref{colour-flavour}, our $f_1$ form factors are given by those in Eq.~\eqref{ourFFs-zero} multiplied
with $c_{f_1}/(c_{a_1}+c_{f_1}+c_{f_1'})$, which yields $\Gamma_{\gamma \gamma} = 3.9(3)$ keV.
}

 We should emphasize that the internal consistency of our calculations is crucial for the quality of our results.
 If any of the three ingredients (meson Bethe-Salpeter amplitude, quark propagator, quark-photon vertex) 
 is replaced with model input,
 the magnitudes and shapes of the form factors can change substantially and $F_2^A$ and $F_3^A$ can even switch signs, 
 see Fig.~\ref{fig:fax-models} in App.~\ref{sec:gi}. In addition, with model inputs
 also electromagnetic gauge invariance is not automatic and the form factors $F_5^A$ and $F_6^A$ are non-zero and sizeable,
 even if the electromagnetic Ward-Takahashi identity (WTI) that relates the quark propagator and quark-photon vertex is intact.
 The rainbow-ladder truncation, on the other hand, satisfies the WTI already at the level of the BSE kernel and thereby automatically 
 ensures the consistency of all ingredients and the electromagnetic gauge invariance of the matrix elements. 
 A detailed discussion of this issue can be found in App.~\ref{sec:gi}.

  In practice, a robust extraction of the form factors for large values of $\eta_+$ is currently limited to $\eta_+ \lesssim 150$ GeV$^2$ near the symmetric limit $|\omega|/\eta_+ \ll 1$.
 For larger values of $|\omega|/\eta_+$ one would  need contour deformations to deal with the singularities in the integrand~\cite{Weil:2017knt}.
 For larger values of $\eta_+$, the results for $F_2^A$ and $F_3^A$ are numerically too noisy because their respective tensors vanish in the symmetric limit.
 For large $\eta_+$ also the angular dependence of the quark-photon vertex dressing functions (which we  neglect here)
 becomes relevant since in the symmetric limit the vertex is tested in kinematics where one quark is soft and the 
 other hard. This would require a moving-frame solution of the quark-photon vertex BSE which is beyond the scope 
 of the present work.

 In the following we provide parametrizations {of the numerical data} of our form factors for further use in the HLbL amplitude. 
 Although for the muon $g-2$ calculation only their behavior at low $Q^2 \ll 10$ GeV$^2$ is relevant, here 
 we also attempt to implement their large $Q^2 \gg 100$ GeV$^2$ behavior. 
 A light-cone expansion entails the following asymptotic behavior
 for $\eta_+ \to \infty$~\cite{Hoferichter:2020lap}: 
 \begin{equation}\label{large-Q2-ax}
    F_1^A \to \frac{f^A_\text{eff}\,m_A^3}{\eta_+^2}\,h_1(w)\,, \quad
    F_2^A \to \frac{2f^A_\text{eff}\,m_A^5}{5\,\eta_+^3}\,h_2(w)\,,
 \end{equation}
 with $w = \omega/\eta_+$ and
 \begin{equation}\label{FA-log}
 \begin{split}
    h_1(w) &=  -\frac{3}{w^2} \left( 1 + \frac{1}{2w}\ln \frac{1-w}{1+w}\right), \\ 
    h_2(w) &=  -\frac{15}{8w^4} \left( 6 + \frac{3-w^2}{w} \,\ln \frac{1-w}{1+w}\right). 
 \end{split}
 \end{equation}  
 In the symmetric limit this implies $h_1(0) = h_2(0) = 1$, whereas in the asymmetric limits
 the $h_i$ diverge logarithmically. 
 The remaining form factor behaves like $F_3^A \sim 1/\eta_+^3$ for large $\eta_+$. 
  The effective decay constant $f^A_\text{eff}$ in Eq.~\eqref{large-Q2-ax} sums over the $a_1$, $f_1$ and $f_1'$.
  Assuming that all three states have the same decay constant $f_A$, then for 
 ideal mixing one obtains
 \begin{equation}\label{feff}
 \begin{split}
    f^A_\text{eff} &\approx \frac{4}{\sqrt{6}}\,(c_{a_1} + c_{f_1} + c_{f_1'}) \,f_A \,.
 \end{split}
 \end{equation} 
  
  In the accessible $\eta_+$ range near the symmetric limit we see the onset of the asymptotic {scaling} behavior $F_1^A \sim 1/\eta_+^2$ and  $F_2^A \sim 1/\eta_+^3$. % as given by our fits.
 $F_3^A$, on the other hand, develops a zero crossing around $\eta_+ \approx 30$ GeV$^2$ and remains positive for large $\eta_+$. 
 
      \begin{table}
    \centering
    \begin{tabular}{l @{\quad} c @{\quad} r @{\quad} c @{\quad} r @{\quad} r @{\quad} c @{\quad} c @{\quad} c} \hline\noalign{\smallskip}
     $i$       & $a_i$       & $b_i$      & $c_i$  & $d_i$ & $e_i$   & $\nu_i$  & $\mu_i$ & $\lambda_i$         \\ \noalign{\smallskip}\hline\noalign{\smallskip}
    $1$        & 1.80\,(7)       & 3.28       & 4.64\,(50)   & 4.74 & 24.9    &  2 & 1.6 &  4.0  \\[0.5mm]
    $2$        & 2.08\,(5)       & 6.45       & 4.74\,(23)   & 8.22 & 11.7    & 3  & 1.6 & 5.0        \\[0.5mm]
    $3$        & $-0.41(1)$      & $-1.31$    & 1.22\,(23)   & $-56.0$ & 64.4    & 3  &   $-$ &   $-$            \\\noalign{\smallskip}\hline
    \end{tabular}
    \caption{Fit parameters for our axial-vector transition form factors $F_i^A(x,w)$.}
    \label{tab:fitpar}
    \end{table}
 
 To match with the asymptotic constraints away from the symmetric limit, we parametrize our form factors for $x \geq 0$ and $|w| < 1$ as follows:
 \begin{equation}\label{eq-fit}
     F_i^A(x,w) =G_i^A(x)\,H_i^A(x,w)
 \end{equation}
 with $x = \eta_+/m_\rho^2$, $m_\rho = 0.77$ GeV, and
 \begin{align}\label{eq:sums}
    G_i^A(x) &= \frac{1}{(1+x)^{2\nu_i}}\left( a_i + b_i\,x + c_i \,x^{\nu_i} \,\frac{x+d_i}{x+e_i}\right), \nonumber \\
    H_1^A(x,w) &= 1 + \sum_{n=1}^N \frac{x^{\mu_1}}{x^{\mu_1}+\lambda_1 n}\,\frac{w^{2n}}{1+\frac{2n}{3}} \,, \\
    H_2^A(x,w) &= 1 + \sum_{n=1}^N \frac{x^{\mu_2}}{x^{\mu_2}+\lambda_2 n}\,\frac{(1+n)\,w^{2n}}{\left(1+\frac{2n}{3}\right)\left(1+\frac{2n}{5}\right)} \,,   \nonumber \\
    H_3^A(x,w) &= 1\,. \nonumber
 \end{align}
 Because of $H_i^A(x,0) = 1$  the form factors in the symmetric limit reduce to the $G_i^A(x)$, 
 whose parameters are listed in Table~\ref{tab:fitpar}.
 From the series expansion of the logarithms in Eq.~\eqref{FA-log} one obtains
 \begin{equation}
 \begin{split}
    h_1(w) &= 1 + \sum_{n=1}^\infty \frac{w^{2n}}{1+\frac{2n}{3}}\,, \\
    h_2(w) &= 1 + \sum_{n=1}^\infty \frac{(1+n)\,w^{2n}}{\left(1+\frac{2n}{3}\right)\left(1+\frac{2n}{5}\right)}\,,
 \end{split}
 \end{equation} 
 which entails that $H_i^A(x\to\infty,w) = h_i(w)$ for $N\to \infty$.
 Therefore, the functions $H_i^A(x,w)$ interpolate
 between $H_i^A(x=0,w) = 1$ and $H_i^A(x\to\infty,w) = h_i(w)$.
 In the absence of asymptotic constraints for $F_3^A$ we set $H_3^A(x,w) = 1$.
 
 {The quality of our parametrisations may be inferred from the comparison in Appendix \ref{sec:num};
 the numerical results and fits are indistinguishable by eye. We note that the convergence behaviour of the sums in 
 Eq.~(\ref{eq:sums}) is rather slow. We used $N=100$ terms to obtain good fits with sub-permille accuracy in the region of low $\eta_+$ shown in Fig.~\ref{fig:fax-full},
 where the singly-virtual case is directly calculable. For larger $\eta_+$ this number would need to further increase to reproduce  the asymptotic logarithms
 via the series expansion.}
 Note also that we do not attempt to provide a realistic parametrization for timelike values in terms of vector-meson poles,
 although in principle this could also be achieved using a series representation. 
 {For the calculation of $a_\mu$, we increased $N$ until our numerical results for the various
 contributions to $a_\mu$ did not change on the sub-permille level. This goal was achieved for $N\approx 60$ and we then
did all calculations with $N=100$ to be on the safe side.}

      \begin{table}[!b]
    \centering
    \begin{tabular}{l @{\quad} c @{\quad} c @{\quad} c @{\quad} r @{\quad} r @{\quad} c @{\quad} c @{\quad} c} \hline\noalign{\smallskip}
     $i$       & $a_i$       & $b_i$      & $c_i$  & $d_i$ & $e_i$   & $\nu_i$  & $\mu_i$ & $\lambda_i$         \\ \noalign{\smallskip}\hline\noalign{\smallskip}
    $1$        & 0.73\,(2)       & 0       & 0.19\,(2)   & $-15.3$ & 14.9    &  1 & 1.6 &  2.6 \\[0.5mm]
    $2$        & 0.41\,(9)       & 0.46       & 0.37\,(5)   & $-10.7$ & 12.4    & 2  & 2.4 & 0.7           \\\noalign{\smallskip}\hline
    \end{tabular}
    \caption{Fit parameters for our scalar transition form factors $F_i^S(x,w)$.}
    \label{tab:fitparS}
    \end{table}

             \begin{figure}[t]
                    \begin{center}
                    \includegraphics[width=0.9\columnwidth]{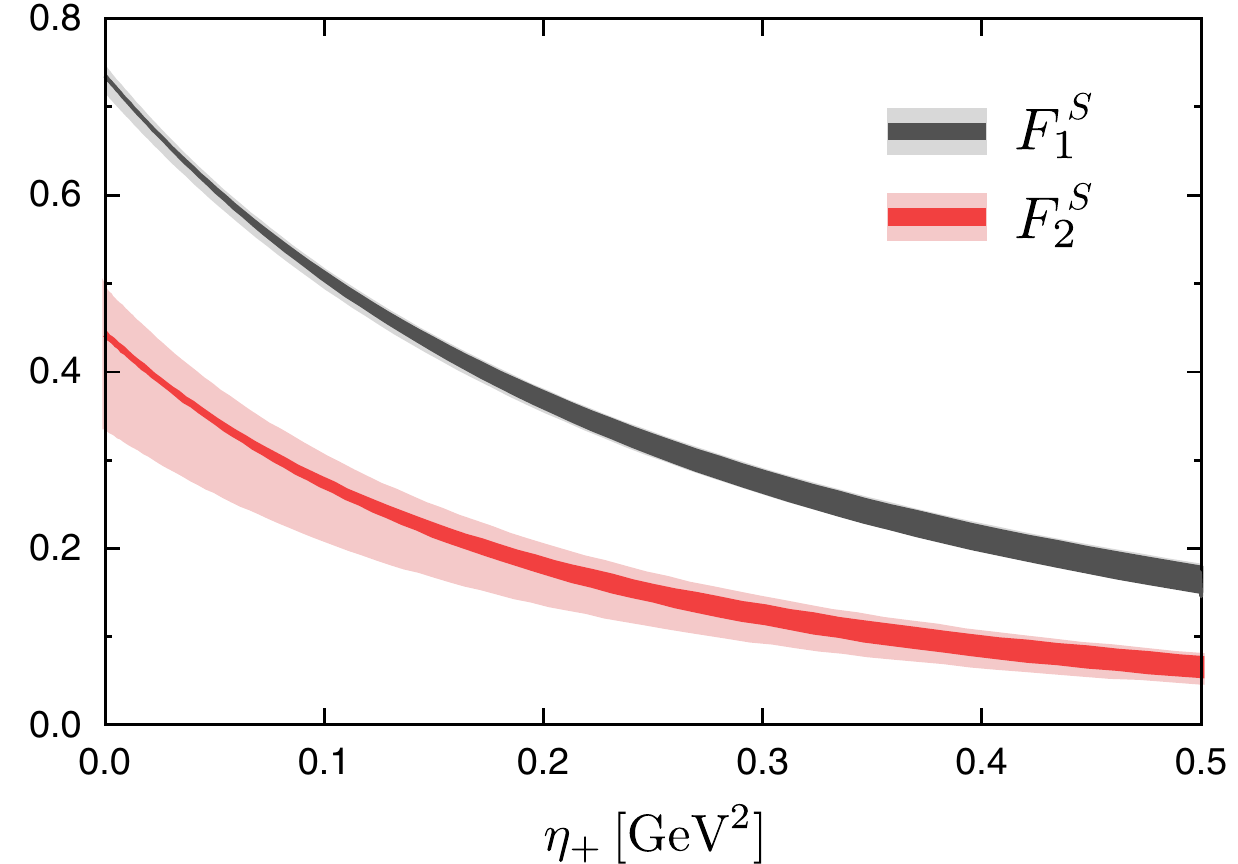}
                    \caption{Scalar transition form factors as functions of $\eta_+$, see Fig.~\ref{fig:fax-full} for details.}\label{fig:fsc-full}
                    \end{center}
            \end{figure} 
 
  \subsection{Scalar transition form factors}\label{sec:scTFF}

  The analogous results for the scalar transition form factors obtained from Eq.~\eqref{mic-decomp},
  $F_i^S(Q^2,{Q'}^2)$ with $i=1,2$, are displayed in Fig.~\ref{fig:fsc-full}.
  Like in the axial-vector case, we sum up the three light flavour states $a_0$, $f_0$ and $f_0'$
  whose colour-flavour factors are given by Eq.~\eqref{colour-flavour}.
  Also in this case the angular dependence is small and yields a narrow band in $\eta_+$ (dark bands).
  The light coloured bands correspond again to the variation of the shape parameter in the effective quark-gluon interaction.
  
  At asymptotically large values $\eta_+ \to \infty$, the two form factors behave as~\cite{Danilkin:2021icn,Hoferichter:2020lap}
 \begin{equation}\label{large-Q2-sc}
    F_1^S \to \frac{3f^S_\text{eff}\,m_S}{5\eta_+}\,h_0(w)\,, \quad
    F_2^S \to \frac{3f^S_\text{eff}\,m_S^3}{5\,\eta_+^2}\,h_0(w)\,,
 \end{equation}  
 again with $w = \omega/\eta_+$ and
 \begin{equation}\label{FAs-log}
 \begin{split}
    h_0(w) &=  \frac{5}{2w^4} \left( 3 - 2w^2 + 3\,\frac{1-w^2}{2w}\ln \frac{1-w}{1+w}\right)  \\ 
           &= 1 + \sum_{n=1}^\infty \frac{w^{2n}}{\left(1+\frac{2n}{3}\right)\left(1+\frac{2n}{5}\right)}\,.
 \end{split}
 \end{equation}  
 The effective decay constant $f^S_\text{eff}$ is analogous to its axial-vector counterpart~\eqref{feff}.
 We parametrize our form factors for $x \geq 0$ and $|w| < 1$ in the same way as before,
 \begin{equation}
     F_i^S(x,w) =G_i^S(x)\,H_i^S(x,w)
 \end{equation}
 with $x = \eta_+/m_\rho^2$, $m_\rho = 0.77$ GeV, and
 \begin{align}
    G_i^S(x) &= \frac{1}{(1+x)^{2\nu_i}}\left( a_i + b_i\,x + c_i \,x^{\nu_i} \,\frac{x+d_i}{x+e_i}\right), \nonumber \\
    H_i^S(x,w) &= 1 + \sum_{n=1}^N \frac{x^{\mu_i}}{x^{\mu_i}+\lambda_i n}\,\frac{w^{2n}}{\left(1+\frac{2n}{3}\right)\left(1+\frac{2n}{5}\right)}  
 \end{align}
 for $i=1,2$.
 The fit parameters in this case are listed in Table~\ref{tab:fitparS}.
 At low $\eta_+$, both form factors fall off considerably faster compared to their asymptotic powers
 and $F_2^S$ even becomes negative at intermediate values of $\eta_+$. 
 At large $\eta_+$ both form factors are  positive and the asymptotic constraint
 $F_2^S/F_1^S \to m_S^2/\eta_+$ from Eq.~\eqref{large-Q2-sc} is reproduced.
 
 Like in the axial-vector case, the consistency of our truncation is essential for these results. 
 If any of the  ingredients is replaced with models, electromagnetic gauge invariance is broken,
 the unphysical form factors $F_{3,4,5}^S$ in Eq.~\eqref{amp-final-sc} are nonzero and 
 the magnitudes and shapes of the physical form factors can differ substantially, as shown in Fig.~\ref{fig:fsc-models} in App.~\ref{sec:gi}.

\section{Axial-vector and scalar meson pole contributions to HLbL}\label{HLbL}

We now proceed to discuss our results for the contribution of axial-vector and scalar states to HLbL in narrow width
approximation. To this end we employ the formalism developed in Refs.~\cite{Colangelo:2017fiz,Hoferichter:2024fsj}. Denoting
the HLbL tensor by
\begin{equation}
\Pi^{\mu\nu\lambda\sigma} = \sum_{i=1}^{54} T_i^{\mu\nu\lambda\sigma} \,\Pi_i\,, 
\end{equation}
with scalar functions $\Pi_i$ and Lorenz structures as in Ref.~\cite{Colangelo:2017fiz}, the HLbL master formula 
reads 
\begin{align}
	a_\mu^\text{HLbL} =& \frac{\alpha^3}{432\pi^2} \int_0^\infty d\Sigma\, \Sigma^3 \int_0^1 dr\, r\sqrt{1-r^2} \int_0^{2\pi} d\phi \,\nonumber\\
	&\times \sum_{i=1}^{12} T_i(\Sigma,r,\phi) \,\bar\Pi_i(q_1^2,q_2^2,q_3^2)\,,
\end{align}
where the Euclidean momenta $Q_{1,2,3}^2=-q_{1,2,3}^2$ are parametrized via \cite{Eichmann:2015nra} 
\begin{align}
Q_1^2 &= \frac{\Sigma}{3} \left( 1 - \frac{r}{2} \cos\phi - \frac{r}{2}\sqrt{3} \sin\phi \right), \notag\\
Q_2^2 &= \frac{\Sigma}{3} \left( 1 - \frac{r}{2} \cos\phi + \frac{r}{2}\sqrt{3} \sin\phi \right),\notag \\
Q_3^2 &= \frac{\Sigma}{3} \left( 1 + r \cos\phi \right).
\end{align}
The kernel functions $T_i$ are given in the appendix of Ref.~\cite{Colangelo:2017fiz}. The twelve scalar functions
$\bar{\Pi}_i$ are generated by six representative scalar functions $\hat{\Pi}_i, i \in \{1,4,7,17,39,54\}$ by
explicit and crossing relations given in Eqs.~(2.7) and (2.8) in Ref.~\cite{Hoferichter:2024fsj}. For a given
contribution, these depend explicitly on the transition form factors and propagators of the respective mesons
as detailed below. 

\subsection{Axial-vector mesons: $J^{PC}=1^{++}$}

In Ref.~\cite{Hoferichter:2024fsj} the six representative scalar functions for the case of axial-vector meson
exchange are given in their Eq.~(4.11), again with momenta $q^2_i=-Q_i^2$ and the axial-vector transition form 
factors $\F^A_1,\F^A_2,\F^A_3$ written in a Minkowski basis which translates into our Euclidean one according to
Eq.~(\ref{hoferichter-comparison}).
These form factors are related to ours via
\begin{equation}\label{hoferichter-comparison-2}
\begin{split}
\F_1^A &= - \frac{\omega}{m_A^2}\,F^A_3\,, \\ 
\F_2^A &= \frac{1}{2} \left( F^A_1 - \frac{\omega}{m_A^2}\,F^A_2  \right) ,\\ 
\F_3^A &= -\frac{1}{2} \left(F^A_1 + \frac{\omega}{m_A^2}\,F^A_2  \right).
\end{split}
\end{equation}

As already mentioned in the introduction, the important issue of short-distance constraints (SDCs) to the HLbL tensor
is tightly connected to the anomaly and the role of axial-vector mesons. Detailed discussions can be found in the
literature \cite{Colangelo:2019lpu,Colangelo:2019uex,Melnikov:2019xkq,Leutgeb:2019gbz,Cappiello:2019hwh,
Ludtke:2020moa,Leutgeb:2021mpu,Colangelo:2021nkr,Leutgeb:2022lqw,Hoferichter:2023tgp,Ludtke:2024ase}. 
Within the framework of holographic QCD, 
it has been shown analytically that a resummation of contributions from an infinite tower of axial-vector states 
indeed serves to satisfy the SDCs \cite{Leutgeb:2019gbz,Cappiello:2019hwh,Leutgeb:2021mpu,Leutgeb:2022lqw}. 
The mechanism is quite interesting:
while the asymptotic behaviour of each and every single axial-vector contribution to the HLbL tensor falls off too
strongly with asymptotic momentum $Q^2$, the infinite resummation generates an extra power of $Q^2$ that leads
to the correct asymptotics. This mechanism has been summarised and systematically compared to other suggested 
solutions in Ref.~\cite{Colangelo:2021nkr}. 

In our approach, we are not in a position to provide the same analytic proof. As already discussed, our TFFs 
automatically have the correct asymptotic large momentum behaviour, since perturbation theory is ingrained 
into the DSE framework. However, our TFFs are calculated not analytically but numerically with the difficulty 
that additional numerical problems appear at large momenta. As discussed above, advanced methods to handle 
these are available in principle but their implementation to the problem at hand requires much more work.
This is relegated to the future, together with a more detailed study of the SDCs in our approach.
For now we follow a more modest goal. We include not an infinite, but a sufficiently rich tower of 
axial-vector states such that their total contribution to $a_\mu$ saturates. In a second step 
we study the impact of SDCs using the matching prescription with the quark loop suggested in 
Refs.~\cite{Colangelo:2019uex,Colangelo:2021nkr}.

We perform the first of these tasks as follows. In section \ref{sec:constraints} we found our TFFs to 
nicely match the experimental constraints, although the masses of our axial-vector states are too low. 
This motivates the assumption that the axTFFs do not change drastically throughout the whole tower of 
states. Thus we may use our result for the axTFFs of the ground state as proxy also for all other states. 
For the masses of the first two multiplets of $a_1$, $f_1$ and $f'_1$ states
we use the central values of the experimentally seen $a_1$ states, i.e. $m_{a_1(1260)}=1230 (40)$ MeV and 
$m_{a_1(1640)}=1655$ MeV (Breit-Wigner masses) \cite{ParticleDataGroup:2024cfk}.\footnote{Assuming mass 
degeneracy within a given multiplet simplifies 
the calculation of its contribution to HLBL, since the corresponding flavour and charge factors can be 
summed analytically \cite{Leutgeb:2019gbz}, which in our conventions from Eq.~\eqref{colour-flavour} amounts to a factor 
\begin{equation*}
  \quad \frac{c_a^2 + c_f^2 + c_{f'}^2}{(c_{a}+c_{f}+c_{f'})^2} 
  = \frac{2N_c\sum_a\left[\text{tr}(t^a \mathcal{Q}^2)\right]^2}{(c_{a}+c_{f}+c_{f'})^2} = \frac{36}{(8+\sqrt{2})^2}\,,
\end{equation*}
with $\mathcal{Q}=\text{diag}(2/3,-1/3,-1/3)$, $t^a=\lambda^a/2$,
$t^0=\mathds{1}/\sqrt{6}$, Gell-Mann matrices $\lambda^a$, and the sum goes over $a=0,3,8$.}
Since the experimental masses of the corresponding $f_1$ and $f'_1$ states are larger, we therefore provide 
an upper bound for those contributions. In order to get a rough idea 
of the size of contributions from higher multiplets, we solved a simple Schrödinger equation for a 
quark model with Richardson type potential $V(r) = -\alpha/r + b\,r$ with constituent quarks of mass 
$M=350$ MeV and matched the two parameters to the masses of the $a_1(1260)$ $(j=1)$ and $a_1(1640)$ $(j=2)$. 
This is achieved to good accuracy with $\alpha=1.6$ and $b=0.068$ GeV$^2$. The numerical results for the 
masses of the radial excitations with main quantum number $n=j+1>3$ then follow the curve 
\begin{equation}
   M^2(n) = (-0.76 + 1.095\, n + 0.016\, n^2)\, \text{GeV}^2,
\end{equation}
which is Regge-like for small $n$. We use these as masses for our tower of axial-vector multiplets.

{Furthermore, as already indicated at the end of section \ref{sec:micro}, we re-normalise our axialvector form factors
using the two-photon decay width
\begin{align}\label{ax-norm}
	|F^A_1(0,0)|^2 &= \frac{48\, \Gamma_{\gamma \gamma}}{\pi \alpha^2 m_A}
\end{align}
of the $f_1(1285)$ with $\Gamma_{\gamma \gamma} = 3.5 (6)(5)$ keV, as measured by the L3 collaboration \cite{L3:2001cyf}. 
}

Our individual integrated results for the lowest-lying multiplets are displayed in the upper row of Table~\ref{tab:axvector}. 
In our calculation we observe that fifteen multiplets are enough to obtain a converged result on the level 
of $0.01 \times 10^{-11}$. To be on the safe side, we stopped after twenty multiplets, which is also our result for the 
whole infinite tower:
\begin{align} 
{a_\mu^{\text{HLbL}}[\text{AV-tower}] }  & {= 23.8\,(5.4)(4.4)(0.7)    \times 10^{-11}\,.}
\end{align}
{The first error propagates the uncertainty of the input value for the two-photon decay width to our final result.
The second error reflects the model-related uncertainty due to the mismatch of the rainbow-ladder axialvector mass as 
compared to the experimental value: Whereas the global TFF is properly rescaled to match Eq.~(\ref{ax-norm}), the 
relative strength of $F^A_1$ to $F^A_{2,3}$ is still affected by this mismatch. To take this into account, we determine 
the difference between using the (re-normalised) TFFs $F_i^A$ obtained from the prescription in Eq.~\eqref{resc}
%detailed at the end of section \ref{sec:micro} 
and using the unnormalized $(F_i^A)^\text{RL}$ obtained with the rainbow-ladder axialvector mass in the tensor
structures as a measure for the possible spread of results. The error reflecting this difference is about 
equal in magnitude to the one inflicted by the uncertainty of the experimental mass of the $a_1(1260)$. Both have been 
already added in quadrature. The third contribution to our error budget stems from the same 
variation of the model parameter $\eta$ that leads to the light shaded bands in our TFFs, Fig.~\ref{fig:fax-full}. 
Finally, the error introduced by using fits to our numerical results for the TFFs when calculating $a_\mu$ 
(cf. Appendix~\ref{sec:num}) as well as the numerical error of the VEGAS routine used to perform the integrations 
for $a_\mu$ are both negligible by orders of magnitude and therefore not shown.} 

      \begin{table}
	\centering 
	\begin{tabular}{ c @{\;\;} | @{\;\;}c @{\;\;\;} @{\;\;}c @{\;\;\;}}                                         \hline 
		                       &      full                  &   matched                       \rule{0pt}{10pt}\\[0.5mm] \hline
		      $j=1$            &     17.4\,(4.0)(4.0)(0.5)  &                                 \rule{0pt}{10pt}\\[0.5mm] \hline
		      $j \le 2$        &     20.9\,(4.8)(4.3)(0.6)  &                                 \rule{0pt}{10pt}\\[0.5mm] \hline
		      $j \le 10$       &     23.6\,(5.4)(4.4)(0.7)  &                                 \rule{0pt}{10pt}\\[0.5mm] \hline
		      $j \le 20$       &     23.8\,(5.4)(4.4)(0.7)  &  24.8\,(6.1) $\times 10^{-11}$    \rule{0pt}{10pt}\\[0.5mm] \hline
	\end{tabular}
	\caption{  Results for $a_\mu^{AV}$ in units of $10^{-11}$. 
		Shown are integrated results for the ground-state multiplet $(j=1)$ as well as
		two, ten and twenty  multiplets. In the second column we display fully integrated results and
		in the third column those integrated below a matching scale, see text for details.}
	\label{tab:axvector}
\end{table}

\begin{figure}[t]
                    \begin{center}
                    \includegraphics[width=0.85\columnwidth]{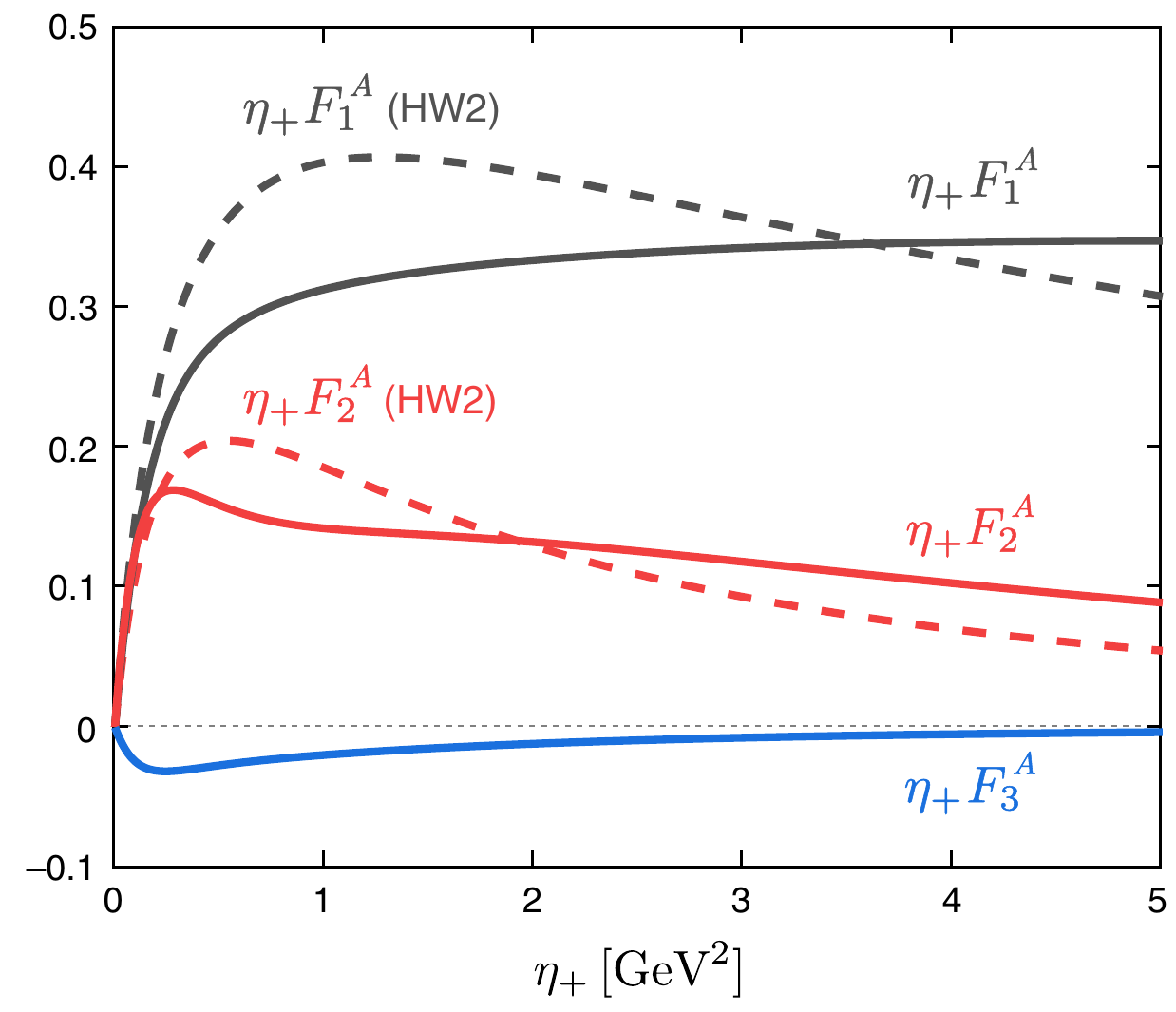}
                    \caption{Comparison of our axial-vector transition form factors in the asymmetric, i.e., singly-virtual limit (solid curves)
                              with the corresponding ones from holographic QCD in the HW2 model~\cite{Leutgeb:2019gbz} (dashed curves; $F_3^A=0$ in their model). All values are in GeV$^2$.}\label{fig:fax-compare}
                    \end{center}
                    \vspace{-4mm}
            \end{figure}

Our result, even when comparing on the level of single axial-vector pole contributions, is considerably 
larger than previous model results such as
Refs.~\cite{Pauk:2014rta,Jegerlehner:2017gek} and the White Paper estimate \cite{Aoyama:2020ynm}. 
It is, however, of comparable size as results from holographic models
\cite{Leutgeb:2019gbz,Cappiello:2019hwh,Leutgeb:2021mpu,Leutgeb:2022lqw,Colangelo:2024xfh}. 
Fig.~\ref{fig:fax-compare} shows a comparison of the axTFFs of our approach with the one of the 
HW2-model of Ref.~\cite{Leutgeb:2019gbz}. Although there are differences in the details of the momentum
dependence, the overall orders of magnitude are similar.

We emphasise again that our framework satisfies constraints from gauge symmetry (cf. App.~\ref{sec:gi}), 
and our representations of the TFFs satisfy the SDCs on the level of the TFFs, Eqs.~(\ref{large-Q2-ax}) 
and (\ref{FA-log}), by construction. To study the impact of the short distance constraints on the hadronic 
light-by-light tensor, we adopt the matching procedure with the perturbative quark loop suggested in
Refs.~\cite{Colangelo:2019uex,Colangelo:2021nkr}. The NLO corrections to
the massless quark loop were determined recently in Ref.~\cite{Bijnens:2021jqo} and found to reduce the error
associated with the quark loop contributions drastically. In Refs.~\cite{Colangelo:2019uex,Colangelo:2021nkr} 
a tower of pseudoscalar meson pole contributions was matched to the quark loop results, but it was pointed out 
that the matching procedure is general and may as well be applied to a tower of axial-vector states. Since there are 
good arguments that this is anyhow what should be done \cite{Leutgeb:2019gbz,Leutgeb:2021mpu}, we adopt the procedure 
here as well. The general idea is to cut the integration of the tower of axial-vector states at a matching scale $Q_\text{min}$
and add the mixed regions with two large and one small momentum by a suppression factor $Q_i^2/(Q_i^2+Q^2_\text{min})$ for
the small momentum $Q_i$ while retaining the cut that the two other momenta are larger than $Q_\text{min}$, see 
\cite{Colangelo:2019uex} for more details. For simplicity and comparability, we choose the same matching scale 
$Q_\text{min}=1.7\,(5)$ GeV as in \cite{Colangelo:2019uex,Colangelo:2021nkr} and therefore obtain the same quark loop 
contribution $a_\mu^{\text{\text{HLbL}}}[\text{q-loop} (N_f=3)] = 4.2\,(1) \times 10^{-11}$ as in Ref.~\cite{Colangelo:2021nkr}.
The reduced contributions from the axial-vector states with this procedure can be found in the lower line of Table 
\ref{tab:axvector}. It is quite satisfactory to see that the overall reduction of these contributions is of the same
order as the quark-loop contributions. We therefore obtain as our final estimate
\begin{align} 
	{a_\mu^{\text{HLbL}}[\text{AV-tower+SDC}]  } & {= [ 20.6_{-2.8}^{+1.3}(4.7)(3.4)(0.3)_{\text{AV-poles}} } \nonumber\\
                                               & {\hspace{3mm}+ 4.2\,(0.1)_{\text{q-loop} (N_f=3)} ] \times 10^{-11}, }\nonumber\\
                                               & {= 24.8\,(6.1) \times 10^{-11} }
\end{align}
for the sum of both axial-vector tower and SDC contributions. {Here, the first error in the result for the AV poles 
reflects the uncertainties in the matching scale $Q_\text{min}$, whereas the second, third and fourth errors reflect
uncertainties from the two-photon width, the variation of model parameters and the uncertainties due to the mass scale, 
similar as above.} For the final result we average the first error and add the other errors in quadrature. 
Our final result is still in the ballpark of results from the holographic 
approach \cite{Leutgeb:2019gbz,Cappiello:2019hwh,Leutgeb:2021mpu,Leutgeb:2022lqw}.

\subsection{Light and heavy scalar mesons: $J^{PC}=0^{++}$}

The six representative scalar functions for the case of scalar meson exchange are given in Eq.~(4.11) of Ref.~\cite{Hoferichter:2024fsj},
again with momenta $q^2_i=-Q_i^2$ and their scalar transition form factors $\F^S_1,\F^S_2$ denoted in a basis which translates 
into ours according to
\begin{align}
F^S_1 = \F^S_1; \hspace*{5mm} F^S_2 = -\F^S_2 \,.
\end{align} 

We start with the light scalar mesons. The $f_0(500)$ has been treated rigorously in the dispersive approach in terms of 
s-wave $\pi\pi$ rescattering, resulting in 
$a_\mu[f_0(500)] = -9 \,(1)  \times 10^{-11}$ \cite{Colangelo:2017qdm,Colangelo:2017fiz}. This result has been confirmed later 
in Ref.~\cite{Danilkin:2021icn}. In the same work, the rescattering contribution of the $f_0(980)$ has been estimated to result in
$a_\mu[f_0(980)](\text{rescattering}) = -0.2 \,(1)  \times 10^{-11}$. In the same reference, a corresponding quark model result in 
narrow width approximation (NWA) has been found to be $a_\mu[f_0(980)](\text{NWA}) = -0.37 \,(6)  \times 10^{-11}$ using 
the normalisation condition
\begin{align}\label{scalar-norm}
|F^S_1(0,0)|^2 &= \frac{4\, \Gamma_{\gamma \gamma}}{\pi \alpha^2 m_S}
\end{align}
with $\Gamma_{\gamma \gamma}[f_0(980)]=0.31\,(5)$ keV \cite{ParticleDataGroup:2024cfk} for the TFFs. 
In order to make contact with their result, we adopt the same normalisation condition for our microscopically calculated TFFs. 
For the mass we use $m_{f_0(980)} = 990(20)$ MeV \cite{ParticleDataGroup:2024cfk}.
We then obtain
\begin{align} 
a_\mu^{\text{HLbL}}[f_0(980)]   &= -0.287 \,(46)\,{(26)}\,(4)  \times 10^{-11}\,,
\end{align}
where the first error is due to the uncertainty in the decay width, the second one {is by far dominated by the uncertainty 
in the experimental mass combined with the much smaller uncertainty due to the ambiguity in the mass factors of 
	the matrix element (cf. corresponding explanations in the axialvector case)},  
and the third one is our systematic model uncertainty {as reflected in the error bands of the TFFs}.
Our result agrees with the rescattering result of Ref.~\cite{Danilkin:2021icn} within error bars. In a similar spirit, 
for the $a_0(980)$ we also adopt the corresponding normalisation condition, this time with 
$\Gamma_{\gamma \gamma}[a_0(980)]=0.5_{-0.1}^{+0.2}$ keV \cite{Lu:2020qeo} and $m_{a_0(980)} = 980(20)$ MeV 
\cite{ParticleDataGroup:2024cfk}:
\begin{align} 
a_\mu^{\text{HLbL}}[a_0(980)]   &= -0.484 \,(145)\,{(45)}\,(8)  \times 10^{-11},
\end{align}
where we averaged over the uncertainties in the decay with. Note that this uncertainty is much larger than the systematic 
uncertainty of our model. Within error bars, our result agrees with the (narrow width) quark model result of 
Ref.~\cite{Danilkin:2021icn} and the very recent dispersive result 
$a_\mu^{\text{HLbL}}[a_0(980)]_{\text{resc.}}   = -0.43(1)  \times 10^{-11}$ \cite{Deineka:2024mzt} taking into account 
re-scattering effects.

For the heavy meson multiplet above 1 GeV the experimental situation is worse. Following \cite{Danilkin:2021icn} we 
use again Eq.~(\ref{scalar-norm}) with $\Gamma_{\gamma \gamma}[f_0(1370)] \leq 3.0^{+1.4}_{-0.9}$ keV and 
$\Gamma_{\gamma \gamma}[a_0(1450)] = 1.05^{+0.5}_{-0.3}$ keV \cite{Lu:2020qeo}. The mass of the $f_0(1370)$ is not very 
well determined experimentally. We use again the Breit-Wigner mass $m_{f_0(1370)} = 1350 (150)$ MeV  \cite{ParticleDataGroup:2024cfk}
with extremely large error. The situation is much better for the $a_0$: $m_{a_0(1450)} = 1439(34)$ MeV. We then arrive at 
\begin{align} 
a_\mu^{\text{HLbL}}[f_0(1370)]  &= -0.673 \,(258)\,{(377)}\,(7)  \times 10^{-11}.\\
a_\mu^{\text{HLbL}}[a_0(1450)]  &= -0.175 \,(66)\,{(22)}\,(2)  \times 10^{-11}.
\end{align}
where we averaged separately over the uncertainties in the decay width and the mass.
These results are in the ballpark of the quark model narrow width estimate of Ref.~\cite{Danilkin:2021icn}.

\section{Summary and conclusions}\label{summary}
\begin{table}
	\centering
	\begin{tabular}{
			c @{\quad}                 c @{\quad}          c     } \hline\noalign{\smallskip}
		                                &  DSE/BSE          & WP   \\ \noalign{\smallskip}\hline\noalign{\smallskip}
	    $\pi$ exchange                  & 62.6\,(1)(1.3)     & 63.0$^{+2.7}_{-2.1}$                 \\ \noalign{\smallskip}
	    $\eta$ exchange                 & 15.8\,(2)(3)(1.0)  & 16.3\,(1.4)                         \\ \noalign{\smallskip}
	    $\eta'$ exchange                & 13.3\,(4)(3)(6)   & 14.5\,(1.9)                         \\ \noalign{\smallskip}\hline\noalign{\smallskip}
		$\pi_0, \eta,\eta'$ exchange    & 91.6\,(1.9)        & 93.8$^{+4.0}_{-3.6}$                 \\ \noalign{\smallskip}\hline\hline\noalign{\smallskip}
        $\pi$ box                       &$-$15.7\,(2)(3)    &$-$15.9\,(2)                        \\ \noalign{\smallskip}
        $K$   box                       &$-$0.48\,(2)(4)    &$-$0.46\,(2)                        \\ \noalign{\smallskip}\hline\noalign{\smallskip}
		$\pi$, $K$ boxes/loops    		&$-$16.2\,(5)       &$-$16.4\,(5)                        \\ \noalign{\smallskip}\hline\hline\noalign{\smallskip}
		s-wave $\pi\pi$ rescattering	&\text{not calc.}   &$-$8.0\,(1.0)                        \\ \noalign{\smallskip}
        higher scalar exchange         	&$-$1.6\,(5)        &$-$2.0\,(2.0)                        \\ \noalign{\smallskip}
		AV exchange (single)      		& { 17.4\,(6.0) }       &   6.0\,(6.0)                      \\ \noalign{\smallskip}
		AV exchange (tower) + SDC   	& { 24.8\,(6.1) }       &  21.0\,(16.0)                     \\[1mm]\hline
	\end{tabular}
	\caption{Results for various contributions to $a_\mu^\text{HLbL}$ in units of $10^{-11}$. Compared are our results
		of Ref.~\cite{Eichmann:2019tjk,Eichmann:2019bqf} (pseudoscalar exchange and pseudoscalar boxes) and this work 
		(axial-vector and scalar exchange) with the White Paper (WP) results of Ref.~\cite{Aoyama:2020ynm}.} 
	\label{tab:summary}
\end{table}
 
In this work we determined axial-vector and scalar meson exchange contributions to the HLbL tensor of $a_\mu$. We worked within
a microscopic approach to QCD using (a truncated set of) Dyson-Schwinger and Bethe-Salpeter equations, which have been successfully 
employed before to determine pseudoscalar exchange and box contributions to HLbL. From an effective underlying 
quark-gluon interaction we determined quark propagators, meson Bethe-Salpeter amplitudes, the quark-photon vertex and subsequently
axial-vector and scalar transition form factors. We demonstrated explicitly that our approach preserves gauge invariance on a microscopic
level and is therefore guaranteed to be free of artefacts stemming from otherwise non-zero extra tensor structures (cf. App.~\ref{sec:gi}). 
Our results for the scalar and axial-vector TFFs complement the ones for the pseudoscalar electromagnetic and transition form factors
determined in Refs.~\cite{Eichmann:2019tjk,Eichmann:2019bqf}. We provided explicit representations of these form factors
that satisfy the large-momentum  constraints. Our pseudoscalar and axial-vector TFFs agree with experimental information at zero 
momenta, see \cite{Eichmann:2019tjk,Eichmann:2019bqf} and the discussion around Eq~(\ref{ffs-exp}) in this work. For the scalar TFFs 
we used experimental information for the two-photon decay width as input to provide for a correct normalisation at zero momentum transfer.
A further caveat concerns the 
short-distance constraints (SDCs) on the HLbL tensor itself. Although we took a tower of axial-vector contributions into account,
we cannot show at present whether an infinite resummation of such a tower would satisfy these SDCs as it does in holographic
QCD \cite{Leutgeb:2019gbz,Cappiello:2019hwh,Leutgeb:2021mpu,Leutgeb:2022lqw}. In contrast to their approach, our TFFs are
given numerically, and much more refined numerical methods would be needed to extract the necessary information. This work
is relegated to the future. Instead, we applied a matching procedure with the perturbative quark loop suggested in
Refs.~\cite{Colangelo:2019uex,Colangelo:2021nkr}, which leads to a combined result of axial-vector and short-distance contributions. 

Our collected results are given in Table~\ref{tab:summary} together with the corresponding results of the first White Paper
\cite{Aoyama:2020ynm}. We find very good agreement on all contributions within error bars. Our central result for the contributions
from a single axial-vector multiplet is about a factor of three larger than the White Paper estimate, whereas the impact of SDCs 
turns out to be smaller than estimated for the White Paper. Taken together, however, the results overlap within error bars. 
To our mind, systematic cross-checks and comparisons between the dispersive approach, 
lattice QCD, holographic QCD and our approach to QCD via functional methods certainly has the potential to further decrease the 
spread of results and consequently the error bar of the axial-vector contributions to HLbL in the future. 

\vspace*{3mm}
{\bf Acknowledgments}\\
We are grateful to Igor Danilkin, Anton Rebhan and Peter Stoffer for discussions and to Anton Rebhan and Peter Stoffer for 
constructive comments on the manuscript.
This work was supported by the Deutsche Forschungsgemeinschaft (DFG) under grant number Fi 970/11-2.

\appendix

   \section{Axial-vector tensor basis}\label{sec:basis}
   
   In this appendix we provide details on the construction of the axial-vector tensor basis in Eq.~\eqref{taui-A}, see also Refs.~\cite{Kuhn:1979bb,Rudenko:2017bel,Roig:2019reh,Hoferichter:2020lap}.
   The basis consists of four transverse tensors $\tau_{1 \dots 4}^{\mu\nu\rho}$  which satisfy the constraints
   \begin{equation}\label{em-gi2}
     Q^\mu \tau_i^{\mu\nu\rho} = 0\,, \qquad
     {Q'}^\nu \tau_i^{\mu\nu\rho} = 0 \,.
   \end{equation}
   The remaining two non-transverse $\tau_{5,6}^{\mu\nu\rho}$ constitute the gauge part whose form factors must vanish.
   In addition, the transversality of the axial-vector meson also enforces   transversality in the meson momentum.
   Eq.~\eqref{taui-A} defines a `minimal' basis in the sense that it is free of kinematic singularities and constraints,
   see e.g. Refs.~\cite{Bardeen:1968ebo,Tarrach:1975tu,Colangelo:2014dfa,Eichmann:2015nra,Eichmann:2018ytt}. Its derivation consists of four steps:
   
   \textbf{Step 1} is to write down the most general linearly independent tensor basis 
   which shares the permutation-group symmetries of the full amplitude. In our case, these are the Bose
   symmetry constraints from Eq.~\eqref{bose}, so all tensors should be Bose-symmetric as well. Furthermore,
   the basis should not contain any kinematic singularities and thus no divisions by the variables $\eta_+$, $\eta_-$, $\omega$ or combinations thereof.
   To this end, we start by writing down all possible expressions of the form $\tau^{\mu\nu\rho}(Q,Q')$ that can be constructed
 from the two momenta $Q$ and $Q'$ and  contain an epsilon tensor:
 \begin{equation}\label{tensors-1}
 \begin{array}{l}
    \varepsilon^{\mu\nu\rho}_Q\,, \\[2mm]
     \varepsilon^{\mu\nu\rho}_{Q'}\,, 
 \end{array}\qquad
 \begin{array}{l}
    \varepsilon^{\mu\nu}_{QQ'} \times \{ Q^\rho, \, {Q'}^{\rho}\}, \\[2mm]
    \varepsilon^{\mu\rho}_{QQ'} \times \{ Q^\nu, \, {Q'}^{\nu}\}, \\[2mm]
    \varepsilon^{\nu\rho}_{QQ'} \times \{ Q^\mu, \, {Q'}^{\mu}\}.
 \end{array}
 \end{equation}
With the averaged four-momentum $\Sigma = (Q+Q')/2$,
the Bose symmetry constraint~\eqref{bose} takes the form
 \begin{equation}
    \Lambda_A^{\mu\nu\rho}(\Sigma,\Delta) \stackrel{!}{=} \Lambda_A^{\nu\mu\rho}(-\Sigma,\Delta)\,.
 \end{equation}
  We now construct linear combinations 
 from Eq.~\eqref{tensors-1} that are either even or odd under this operation.
 Because the variable $\omega = \Sigma\cdot \Delta$ from Eq.~\eqref{vars-1} is also odd (whereas $\eta_\pm$ are even),
 we obtain even tensors if we attach a factor $\omega$ to the odd tensors:
 \begin{equation}\label{tensors-2}
   \begin{split}
      K_1^{\mu\nu\rho} &= \varepsilon^{\mu\nu}_{QQ'} \Delta^\rho\,, \\
      K_2^{\mu\nu\rho} &= \omega\,\varepsilon^{\mu\nu}_{QQ'} \Sigma^\rho\,, \\
      K_3^{\mu\nu\rho} &= \varepsilon^{\mu\nu\rho}_\Sigma\,, \\
      K_4^{\mu\nu\rho} &= \omega\,\varepsilon^{\mu\nu\rho}_\Delta\,, \\
      K_5^{\mu\nu\rho} &= {Q'}^\mu \varepsilon^{\nu\rho}_{QQ'} + Q^\nu \varepsilon^{\mu\rho}_{QQ'}\,, \\
      K_6^{\mu\nu\rho} &= \omega\left({Q'}^\mu \varepsilon^{\nu\rho}_{QQ'} - Q^\nu \varepsilon^{\mu\rho}_{QQ'}\right), \\
      K_7^{\mu\nu\rho} &= Q^\mu \varepsilon^{\nu\rho}_{QQ'} + {Q'}^\nu \varepsilon^{\mu\rho}_{QQ'}\,,\\
      K_8^{\mu\nu\rho} &= \omega\left(Q^\mu \varepsilon^{\nu\rho}_{QQ'} - {Q'}^\nu \varepsilon^{\mu\rho}_{QQ'}\right).
   \end{split}
 \end{equation}
 It turns out that only six tensors are linearly independent, so we can drop $K_7$ and $K_8$:
 \begin{equation}
 \begin{split}
    K_7 &= -K_1 + 2\,(\eta_+-\eta_-)\,K_3 - K_4 + K_5\,, \\
    K_8 &= -2K_2 + 2\omega^2 K_3 - (\eta_++\eta_-)\,K_4 - K_6\,.
 \end{split}
 \end{equation}
 The amplitude can thus be written as
 \begin{equation}\label{amp-k}
    \Lambda^{\mu\nu\rho}_A(Q,Q') = \sum_{i=1}^6 c_i(\eta_+,\omega)\,K_i^{\mu\nu\rho}(Q,Q')\,.
 \end{equation}
 Because all $K_i$ are even and the amplitude itself is even, also the form factors $c_i(\eta_+,\omega)$ must be even and
 so they can only depend on $\omega^2$.

   \textbf{Step 2} is to work out the gauge invariance constraints without introducing kinematic singularities,
   which is referred to as the Bardeen-Tung-Tarrach procedure~\cite{Bardeen:1968ebo,Tarrach:1975tu,Hoferichter:2020lap}.
  Gauge invariance implies that the amplitude must be transverse in the photon legs, Eq.~\eqref{em-gi}.
 The tensors $K_1$ and $K_2$ already satisfy both conditions individually. For the remaining ones we obtain
 \begin{equation} \renewcommand{\arraystretch}{1.4}
 \begin{array}{rl}
    Q^\mu K_3^{\mu\nu\rho} &\!\!\!= \frac{1}{2}\,\varepsilon^{\nu\rho}_{QQ'}\,, \\
    Q^\mu K_4^{\mu\nu\rho} &\!\!\!= -\omega\,\varepsilon^{\nu\rho}_{QQ'}\,, \\
    Q^\mu K_5^{\mu\nu\rho} &\!\!\!= \eta_-\,\varepsilon^{\nu\rho}_{QQ'}\,, \\
    Q^\mu K_6^{\mu\nu\rho} &\!\!\!= \omega \eta_-\,\varepsilon^{\nu\rho}_{QQ'} \,,
 \end{array} \quad
 \begin{array}{rl}
    {Q'}^\nu K_3^{\mu\nu\rho} &\!\!\!= \frac{1}{2}\,\varepsilon^{\mu\rho}_{QQ'}\,, \\
    {Q'}^\nu K_4^{\mu\nu\rho} &\!\!\!= \omega\,\varepsilon^{\mu\rho}_{QQ'}\,, \\
    {Q'}^\nu K_5^{\mu\nu\rho} &\!\!\!= \eta_-\,\varepsilon^{\mu\rho}_{QQ'}\,, \\
    {Q'}^\nu K_6^{\mu\nu\rho} &\!\!\!= -\omega \eta_-\,\varepsilon^{\mu\rho}_{QQ'} \,.
 \end{array}
 \end{equation} 
 The transversality conditions then amount to
 \begin{equation}
 \begin{split}
     c_3 &= 2\left( \omega c_4 - \eta_- (c_5+\omega c_6)\right), \\
     c_3 &= -2\left( \omega c_4 + \eta_- (c_5-\omega c_6)\right),
 \end{split}
 \end{equation}
 and taken together these result in
 \begin{equation}\label{elim}
     c_3 = -2\eta_- c_5\,, \qquad c_4 = \eta_- c_6\,.
 \end{equation}
 Plugging this back into Eq.~\eqref{amp-k} yields
 \begin{equation}
    \Lambda_A^{\mu\nu\rho} = c_5\,X_1^{\mu\nu\rho} + c_6\,X_2^{\mu\nu\rho} + c_2\,X_3^{\mu\nu\rho} + c_1\,X_4^{\mu\nu\rho} \,,
 \end{equation}
 where the four $X_i^{\mu\nu\rho}$  are fully transverse and given by
 \begin{equation}
 \begin{split}
     X_1 &= K_5 - 2\eta_- K_3\,, \\
     X_2 &= K_6 + \eta_- K_4\,,  \\
     X_3 &= K_2\,, \\ 
     X_4 &= K_1\,.
 \end{split}
 \end{equation}
 The transversality of the two new tensors $X_1$ and $X_2$ can be made manifest by writing
 \begin{equation}\label{Xi}
 \begin{split}
    X_1^{\mu\nu\rho} &= \varepsilon^{\mu\rho\alpha}_Q \,t^{\alpha\nu}_{QQ'} - t^{\mu\alpha}_{QQ'}\,\varepsilon^{\nu\rho\alpha}_{Q'}\,, \\
    X_2^{\mu\nu\rho} &= -\omega \left( \varepsilon^{\mu\rho\alpha}_Q \,t^{\alpha\nu}_{QQ'} + t^{\mu\alpha}_{QQ'}\,\varepsilon^{\nu\rho\alpha}_{Q'} \right) \,,\\
 \end{split}
 \end{equation}
 The tensors $X_1 \dots X_4$ carry mass dimensions 3, 5, 5, 3, respectively,
 so we may form the linear combinations
  \begin{equation} 
  \begin{array}{rl}
     \tau_1 &= \displaystyle \frac{X_1-X_4}{2m_A^3}\,, \\[4mm]
     \tau_2 &= \displaystyle \frac{X_2+2X_3}{2m_A^5}\,, 
  \end{array}\qquad
  \begin{array}{rl}
     \tau_3 &= \displaystyle \frac{2X_3}{m_A^5}\,, \\[4mm]
     \tau_4 &= \displaystyle \frac{X_4}{m_A^3}\,. 
  \end{array}
 \end{equation}
 These are the transverse tensors in Eq.~\eqref{taui-A}.

 \textbf{Step 3:} Gauge invariance is not always automatic, e.g. in the presence of a  cutoff or,
    as demonstrated in Sec.~\ref{sec:gi}, in model calculations whose ingredients do not follow
    from a consistent truncation that preserves the vector Ward-Takahashi identity (WTI).
    Projecting such an amplitude onto the four tensors $\tau_{1\dots 4}$ can  induce kinematic singularities
    in the corresponding form factors. Thus one must  complete the basis  by adding those tensors that were eliminated  in Eq.~\eqref{elim}, 
    namely $K_3$ and $K_4$:
 \begin{equation} 
     \tau_5 = \frac{K_3}{m_A}\,, \qquad
     \tau_6 = \frac{K_4}{m_A^3}\,.
 \end{equation}    
 These constitute the gauge part. One thus arrives at a
     complete tensor basis consisting of a transverse part and a gauge part,
   whose associated form factors are now guaranteed to be free of kinematic singularities. 
   For the scalar two-photon amplitude $\Lambda_S^{\mu\nu}$ the analogous procedure can be found in Sec. II of Ref.~\cite{Eichmann:2015nra}.
   Other examples are the quark-photon vertex~\cite{Ball:1980ay},
   which consists of a transverse part and a Ball-Chiu part, where the latter satisfies a WTI,
   or nucleon Compton scattering and nucleon-to-resonance
   transition amplitudes~\cite{Eichmann:2018ytt}.

 \textbf{Step 4:} 
 Because the onshell axial-vector meson is also transverse, technically we have another transversality condition % of the form %we want to work out the transversality in the axial-vector-meson leg. The on
 \begin{equation}\label{cond-transversality-2}
    \Delta^\rho \Lambda_A^{\mu\nu\rho}(Q,Q') = 0 
 \end{equation}
 which can be worked out along the same lines.
 To this end, we start from Eq.~\eqref{taui-A} and keep only the first four tensors
 given that $F_5^A$ and $F_6^A$ vanish by gauge invariance.
 The contraction of $\Delta^\rho$ with the tensors $\tau_i$ in Eq.~\eqref{taui-A} yields
 \begin{equation}
 \begin{split}
     \Delta^\rho\,\tau_1^{\mu\nu\rho} &= -\frac{\eta_+}{m_A^3}\,\varepsilon^{\mu\nu}_{QQ'}\,, \\
     \Delta^\rho\,\tau_2^{\mu\nu\rho} &= \frac{\omega^2}{m_A^5}\,\varepsilon^{\mu\nu}_{QQ'}\,, \\
     \Delta^\rho\,\tau_3^{\mu\nu\rho} &= \frac{2\omega^2}{m_A^5}\,\varepsilon^{\mu\nu}_{QQ'}\,, \\
     \Delta^\rho\,\tau_4^{\mu\nu\rho} &= \frac{\Delta^2}{m_A^3}\,\varepsilon^{\mu\nu}_{QQ'}\,. \\
 \end{split}
 \end{equation}
 Unfortunately, the resulting constraint
 \begin{equation}
    \eta_+\,F_1^A - \frac{\omega^2}{m_A^2}\,(F_2^A+2F_3^A) - \Delta^2 F_4^A = 0\,,
 \end{equation}
 where $\Delta^2 = 2\,(\eta_+-\eta_-)$,
 cannot be solved without introducing kinematic singularities. If we solve for $F_4^A$ and plug the result back into Eq.~\eqref{amp-final}, 
 the resulting amplitude $(\Lambda_A^\perp)^{\mu\nu\rho}$ has a kinematic singularity   at $\Delta^2 = 0$:
 \begin{equation}\label{amp-6}
 \begin{split}
    (\Lambda_A^\perp)^{\mu\nu\rho} = m_A\,\Big[ & F_1^A\left( \tau_1^{\mu\nu\rho} + \frac{\eta_+}{\Delta^2}\,\tau_4^{\mu\nu\rho}\right)  \\
                                             + & F_2^A \left( \tau_2^{\mu\nu\rho} - \frac{\omega^2}{\Delta^2 m_A^2}\,\tau_4^{\mu\nu\rho} \right) \\
                                             + & F_3^A\left( \tau_3^{\mu\nu\rho} - \frac{2\omega^2}{\Delta^2 m_A^2}\,\tau_4^{\mu\nu\rho}\right)\Big]\,.
 \end{split}
 \end{equation}
 Because the onshell axial-vector meson amplitude is defined at $\Delta^2 = -m_A^2$, however, this is of no  concern in practice
 because the limit $\Delta^2 = 0$ is never probed.
 Eq.~\eqref{amp-6} is just the transverse projection of the amplitude:
 \begin{equation}
    (\Lambda_A^\perp)^{\mu\nu\rho} = T^{\rho\lambda}_\Delta \sum_{i=1}^3 F_i^A\,\tau_i^{\mu\nu\lambda}\,, \quad  T^{\rho\lambda} = \delta^{\rho\lambda} - \frac{\Delta^\rho \Delta^\lambda}{\Delta^2}\,.
 \end{equation}
 Eq.~\eqref{cond-transversality-2} does therefore not provide any new information; it  simply eliminates $F_4^A$ while  $F_{1,2,3}^A$ remain unchanged.
 In the microscopic expression~\eqref{mic-decomp},
 the origin of the transverse projector is the axial-vector meson amplitude.
 One may then equivalently evaluate Eq.~\eqref{mic-decomp} without that transverse projector,
 project on all six tensors and finally keep only the three physical form factors.

   \section{Comparison with model calculations}\label{sec:gi}
   
\begin{figure*}[t]
	\begin{center}
		\includegraphics[width=1\textwidth]{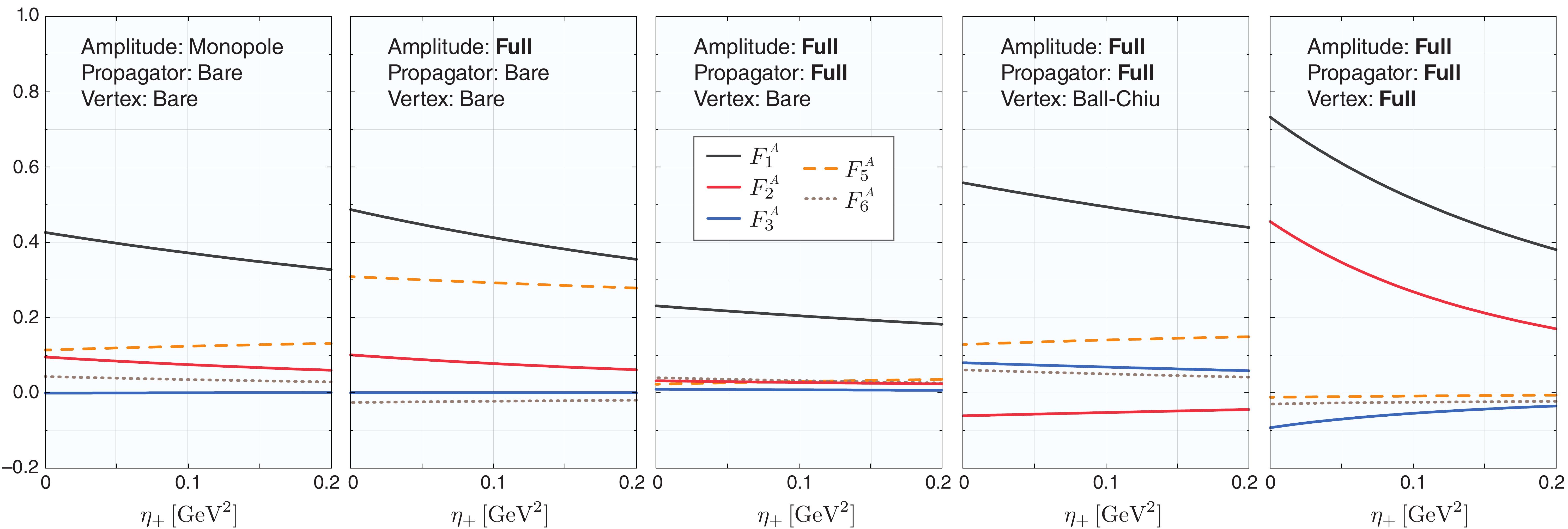}
		\caption{Axial-vector transition form factors obtained from Eq.~\eqref{mic-decomp} using various model inputs.
			The rightmost panel corresponds to the full rainbow-ladder calculation (without the rescaling from Eq.~\eqref{resc}).}\label{fig:fax-models}
	\end{center}
	\begin{center}
		\includegraphics[width=1\textwidth]{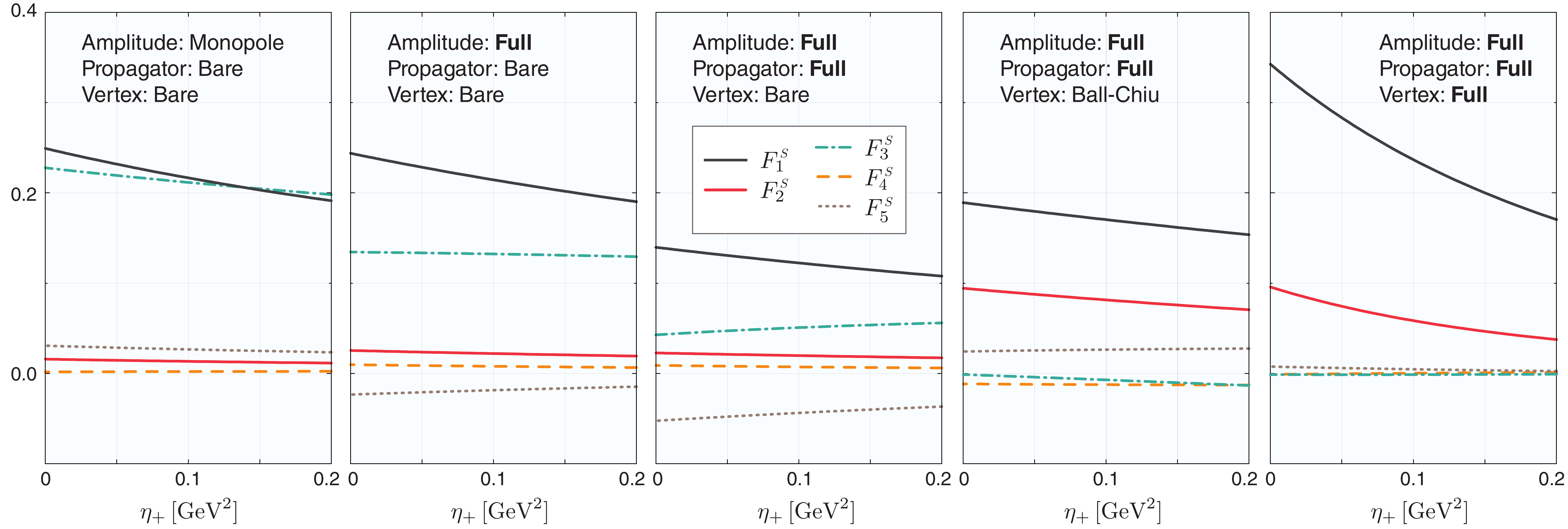}
		\caption{Scalar transition form factors obtained from Eq.~\eqref{mic-decomp} using various model inputs  
			analogous to Fig.~\ref{fig:fax-models}.}\label{fig:fsc-models}
	\end{center}
\end{figure*}

   It is instructive to compare the results in Fig.~\ref{fig:fax-full} with model calculations of the axial-vector meson transition matrix element.
   To do so, we start with the simplest option, namely a tree-level quark-photon vertex $\Gamma^\mu = i\gamma^\mu$, 
   a free quark propagator $S(k) = (-i\slashed{k} + m)/(k^2+m^2)$ with a constituent-quark mass $m$,
   and a monopole Bethe-Salpeter amplitude 
   \begin{equation}\label{monopole-amp}
      \Gamma_A^\rho(k,\Delta) = \frac{\mC m^2}{k^2+m^2}\,i\gamma^\rho \gamma_5\,,
   \end{equation} 
   where $\mC$ is a strength parameter. The resulting form factors in the symmetric limit for $\mC = 30$ and $m=0.5$ GeV  are shown in the leftmost panel of
   Fig.~\ref{fig:fax-models}. 
   %In this case $F_3^A$ vanishes. More importantly, t
   The two form factors $F_5^A$ and $F_6^A$ do not vanish,
   which signals that electromagnetic gauge invariance is broken. This may seem surprising given that the combination of a bare propagator and bare quark-photon vertex
   does satisfy the electromagnetic Ward-Takahashi identity (WTI)
   \begin{equation}\label{wti}
       Q^\mu \,\Gamma^\mu(k,Q) = S^{-1}(k+\tfrac{Q}{2}) - S^{-1}(k-\tfrac{Q}{2}) \,.
   \end{equation}
   In fact, in this simple model the form factors can be calculated analytically using a Feynman parametrization.
   In the limit where all momenta vanish ($\omega = \eta_+ = \eta_- = 0$), and modulo colour and flavour traces, this yields
   \begin{equation}
      F_{1\dots 6}^A = \frac{\mC}{(4\pi)^2} \left\{ \frac{2\beta}{5}\,, \; \frac{8\beta^2}{315}\,, \; 0 \,, \; -\frac{\beta}{3}\,, \; \frac{4}{3}\,, \; -\frac{\beta}{30} \right\} 
   \end{equation}
   with $\beta = m_A^2/m^2$. 
   In this approximation $F_3^A$ vanishes identically for any momentum configuration.
   Moreover, the transversality of the axial-vector meson would eliminate $F_4$, but the remaining form factors
   $F_5^A$ and $F_6^A$ from the gauge part are still nonzero. 
   Our calculation agrees with the analytic result in this limit, which serves as a useful check.
   Had we projected onto the first three tensors alone, the corresponding form factors would have picked up kinematic singularities.
 
   Next, we exchange the model amplitude~\eqref{monopole-amp} by the full amplitude calculated from its BSE (second panel in Fig.~\ref{fig:fax-models}),
   see App.~\ref{sec:RL} for details.
   This does not improve the situation; the spurious form factor $F_5^A$ is now even larger.
   In the third panel we implement the full propagator calculated from the quark DSE
   but keep the vertex bare, which violates the WTI.
   In the fourth panel we exchange the bare vertex by the Ball-Chiu vertex, which is the solution of the WTI~\eqref{wti} for a general quark propagator.
   $F_1^A$ is now much larger, $F_2^A$ switched sign and became negative, and $F_3^A$ is nonzero and positive. $F_5^A$ and $F_6^A$ are again sizeable
   even though the WTI holds.
   
   Finally, the  result including the full quark-photon vertex obtained from its BSE is shown in the fifth panel. 
   The momentum falloff of all three form factors is now considerably steeper, $F_2^A$ is large and positive and $F_3^A$ small and negative.
   Most importantly, the gauge parts are now essentially zero. Their small remainders are presumably due to the fact that we 
   neglect the angular dependence of the quark-photon vertices in the form factor calculation
   to avoid problems with  analytic continuations (which could be cured by solving the vertex BSE in a moving frame).

   Fig.~\ref{fig:fsc-models} shows the same analysis for the scalar transition form factors using various model inputs.
   In the first panel we implement a bare propagator, a bare vertex, and a monopole amplitude
   \begin{equation}\label{monopole-amp-sc}
      \Gamma_S(k,\Delta) = -\frac{\mC m^2}{k^2+m^2}\,\gamma_5\,,
   \end{equation}   
   again with $\mC=30$ and a constituent-quark mass $m=0.5$ GeV. Clearly, the three form factors $F_3^S$, $F_4^S$ and $F_5^S$ from the gauge part are nonzero.
   This does not change when we exchange selected model ingredients with their calculated results (second to fourth panels).
   Only if all ingredients come from the consistent rainbow-ladder calculation, the unphysical form factors  vanish  (fifth panel).

   It is interesting that the
   consistency between the quark propagator and quark-photon vertex through the WTI  in Eq.~\eqref{wti} is sufficient
   to ensure electromagnetic gauge invariance for electromagnetic form factors,
   where it boils down to charge conservation in the limit $Q^2 =0$. For example, the neutron electromagnetic form factor automatically
   satisfies $G_E^n(Q^2=0)=0$ if the WTI is satisfied, irrespective of the form of the baryon amplitude. 
   By constrast, Figs.~\ref{fig:fax-models} and~\ref{fig:fsc-models} show that for two-photon matrix elements electromagnetic gauge invariance
   operates at a deeper level, namely in the truncation of the
   DSEs and BSEs which simultaneously generate the quark propagator, quark-photon vertex and the meson  amplitudes.
   The rainbow-ladder kernel ensures this because it satisfies both the vector and axial-vector WTIs.
   For calculations with model input, however, that link is broken.

\begin{figure*}[t]
	\begin{center}
		\includegraphics[width=0.9\textwidth]{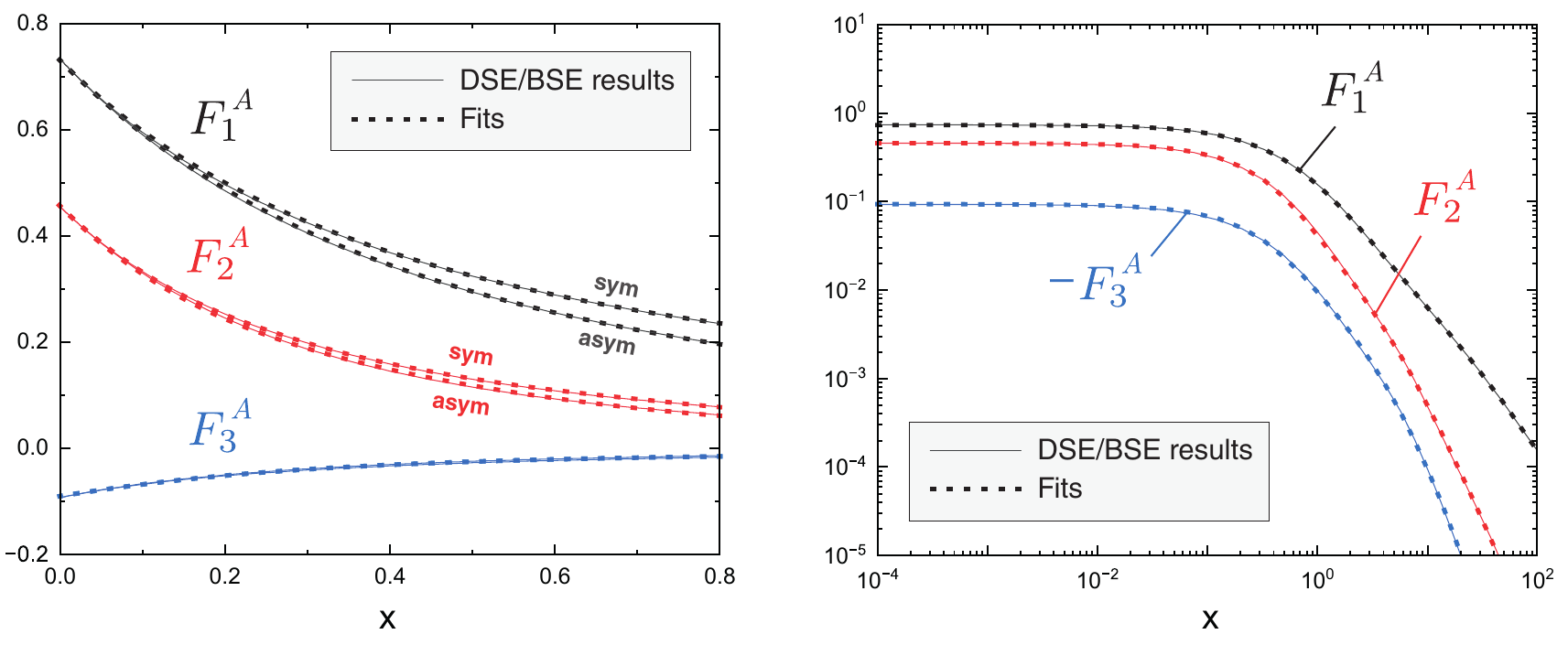}
		\caption{{Axial-vector transition form factors $F_i^A(x,w)$ obtained from our numerical calculation  and compared with 
			our fit functions, Eq.~\eqref{eq-fit}, without the rescaling from Eq.~\eqref{resc}. The left diagram shows the TFFs both in the symmetric ($w=0$) and asymmetric 
			limit ($w=1$) on a linear scale, whereas in the right diagram the TFFs in the symmetric limit are shown on a logarithmic scale.}}
		\label{fig:fit}
	\end{center}
\end{figure*}

   \section{Summary of technical elements}\label{sec:RL}

In the following we briefly outline the various steps needed to calculate the quark propagators, Bethe-Salpeter amplitudes
and the quark-photon vertex needed to determine the axial-vector and scalar TFFs in the functional DSE approach. 
Details can be found in
\cite{Maris:1999bh,Maris:2002mz,Goecke:2010if,Goecke:2012qm,Raya:2015gva,Raya:2016yuj,Eichmann:2017wil,Eichmann:2019tjk}
and the review articles \cite{Maris:2003vk,Maris:2005tt,Eichmann:2016yit}. 

The necessary input to Eq.~\eqref{mic-decomp} is determined from a combination of DSEs and BSEs. The Bethe-Salpeter 
amplitude of a scalar/axial-vector meson and the quark-photon vertex satisfy (in-)homogeneous BSEs
\begin{align}
[\Gamma^{(\rho)}_\text{M}(p,P)]_{\alpha\beta} &= \int_q  [ \mathbf{K}(p,q,P)]_{\alpha\gamma;\delta\beta}   \nonumber \\
& \quad \times [S(q_+)\, \Gamma^{(\rho)}_\text{M}(q,P)\,S(q_-)]_{\gamma\delta}\,,  \label{mesonBSE} \\
[\Gamma^{\mu}(p,P)]_{\alpha\beta} &= Z_2\,i\gamma^\mu_{\alpha\beta} + \int_q  [ \mathbf{K}(p,q,P)]_{\alpha\gamma;\delta\beta} \nonumber  \\
& \quad \times [S(q_+)\, \Gamma^\mu(q,P)\,S(q_-)]_{\gamma\delta}\,,  \label{qqgamma}
\end{align}
where $\mathbf{K}$ is the Bethe-Salpeter kernel, $Z_2$ the quark renormalisation constant, and in both equations 
$q_\pm = q \pm P/2$. The quark propagator $S$ is given by its DSE,
\begin{align}
S^{-1}(p) &= Z_2\,(i\slashed{p} + Z_m m_q) \\
&- Z_{1f} \,g^2 \,C_F\int_q i\gamma^\mu  \, S(q) \, \Gamma^\nu_\text{qg}(q,p)\,D^{\mu \nu}(k) \,,
\end{align}
where $m_q$ is the current-quark mass, $k=q-p$, $C_F=4/3$,
$D^{\mu \nu}$ is the dressed gluon propagator, $\Gamma^\nu_\text{qg}$ the dressed quark-gluon vertex
and $Z_2$, $Z_m$ and $Z_{1f}$ are renormalisation constants.
The gluon propagator and quark-gluon vertex satisfy their own DSEs which include further $n$-point functions, 
so that in all practical applications the tower of DSEs needs to be truncated.

We work in Landau gauge and use the rainbow-ladder truncation, which together with more advanced schemes has 
been reviewed in Ref.~\cite{Eichmann:2016yit}. To this end one defines an effective running coupling $\alpha(k^2)$ 
that incorporates dressing effects of the gluon propagator and the quark-gluon vertex. In the quark DSE this entails
\begin{equation}
\begin{split}
&Z_{1f}\,g^2\,\Gamma^\nu_\text{qg}(q,p)\,D^{\mu \nu}(k) \; \to \; Z_2^2\,\frac{4\pi\alpha(k^2)}{k^2}\,T^{\mu\nu}_k\,i\gamma^\nu \\[-3mm]
&
\end{split}
\end{equation}
with transverse projector $T^{\mu \nu}_k = \delta^{\mu\nu} - k^\mu k^\nu/k^2$.
The kernel $\mathbf{K}$ in the BSEs (\ref{mesonBSE}--\ref{qqgamma}) is uniquely related to the
quark-self energy by vector and axial-vector Ward-Takahashi identities. In rainbow-ladder truncation the kernel is given by
\begin{equation}
[ \mathbf{K}(p,q,P)]_{\alpha\gamma;\delta\beta} \; \to \; Z_2^2\,\frac{4\pi\alpha(k^2)}{k^2}\, i\gamma^\mu_{\alpha\gamma}\,T^{\mu\nu}_k\, i\gamma^\nu_{\delta\beta}\,.
\end{equation}
This construction satisfies chiral constraints such as the Gell-Mann-Oakes-Renner relation and ensures the
(pseudo-)Goldstone boson nature of the pion. Once we have specified the explicit shape of the effective
interaction $\alpha(k^2)$, all elements of the calculation of the form factors
follow and there is no room for any additional adjustments.

Similarly to our previous work on the pion TFF~\cite{Eichmann:2017wil}, the pseudoscalar pole
contributions to HLbL \cite{Eichmann:2019tjk} and the pseudoscalar box contributions to HLbL 
\cite{Eichmann:2019bqf} we use the Maris-Tandy model for the effective coupling $\alpha(k^2)$, 
Eq.~(10) of Ref.~\cite{Maris:1999nt}, with parameters $\Lambda=0.74$ GeV and $\eta = 1.8 \pm 0.2$ 
(the parameters $\omega$ and $D$ therein
are related to the above via $\omega D = \Lambda^3$ and $\omega=\Lambda/\eta$). The scale $\Lambda$ is
fixed via experimental input; we use the pion decay constant for this purpose. The variation of $\eta$
then changes the shape of the quark-gluon interaction at small momenta, cf.~Fig. 3.13 in Ref.~\cite{Eichmann:2016yit},
and we use it as a rough estimate of the truncation error. We work in the isospin symmetric
limit of equal up/down quark masses. With a current light quark
mass of $m_q=3.57$ MeV at a renormalisation point $\mu=19$~GeV we obtain a pion mass and pion decay constant of
$m_{\pi^0} = 135.0(2)$ MeV and $f_{\pi^0} = 92.4(2)$ MeV. With the strange-quark mass fixed at $m_s=85$ MeV
the resulting kaon mass is $m_{K} = 495.0(5)$~MeV. 

{   
   \section{Numerical results for TFFs and fits}\label{sec:num}
In Fig.~\ref{fig:fit} we show exemplary results for our axialvector form factors $F_i^A(x,w)$ in symmetric and asymmetric kinematics on a 
linear scale (left diagram) and for symmetric kinematics on a logarithmic scale (right diagram). The variables $x$ and $w$ have been 
defined in Sec.~\ref{sec:constraints}:
\begin{equation}
   x = \frac{\eta_+}{m_\rho^2} = \frac{Q^2 + {Q'}^2}{2 m_\rho^2}\,, \quad 
   w = \frac{\omega}{\eta_+} = \frac{Q^2 - {Q'}^2}{Q^2 + {Q'}^2}\,,
\end{equation}
where $m_\rho = 0.77$ GeV is also a fit parameter.
The numerical results and fits are indistinguishable by eye, 
thus underlining the high quality of the parametrizations. 
For the scalar TFFs the fit quality is similar.} 

\bibliography{baryonspionff}

\end{document}